\documentclass[rmp,aps, twocolumn, amsmath,amssymb,graphicx,longbibliography]{revtex4-2}

\usepackage{bm, xcolor}
\usepackage{graphicx}
\usepackage{braket}
\usepackage{hyperref}
\usepackage{cleveref}
\usepackage{comment}

\definecolor{pal3}{rgb}{0.5961, 0.3059, 0.6392}

\newcommand{\TTSB}[0]{S$\tau$B}
\begin{document}

\title{{\it \bf Colloquium}: Quantum and Classical Discrete Time Crystals}

\author{Michael P. Zaletel}
    \affiliation{Department of Physics, University of California, Berkeley, California 94720 USA}
    \affiliation{Materials Science Division, Lawrence Berkeley National Laboratory, Berkeley, California 94720 USA}

 \author{Mikhail Lukin}
    \affiliation{Department of Physics, Haarvard University, Cambridge, MA 02138 USA}

 \author{Christopher Monroe}
    \affiliation{Duke Quantum Center, \\ 
    Department of Electrical and 
    Computer Engineering, \\ 
    Department of Physics, \\
    Duke University, \\
    Durham NC 27701  USA}

   \author{Chetan Nayak}
   \affiliation{Microsoft Quantum, Station Q, University of California, Santa Barbara, CA 93106 USA}
    \affiliation{Department of Physics, University of California, Santa Barbara, CA 93106 USA}

    \author{Frank Wilczek}
    \affiliation{Center for Theoretical Physics, MIT, Cambridge MA 02139 USA;}
    \affiliation{T. D. Lee Institute and Wilczek Quantum Center, \\
SJTU, Shanghai;}
\affiliation{Arizona State University, \\ Tempe AZ USA;}
\affiliation{Stockholm University \\ Stockholm, Sweden}

\author{Norman Y. Yao}
    \affiliation{Department of Physics, University of California, Berkeley, California 94720 USA}

\date{\today{}}

\begin{abstract}
The spontaneous breaking of time translation symmetry has led to the discovery of a new phase of matter – the discrete time crystal. Discrete time crystals exhibit rigid subharmonic oscillations, which result from a combination of many-body interactions, collective synchronization, and ergodicity breaking. This Colloquium reviews recent theoretical and experimental advances in the study of quantum and classical discrete time crystals. 
We focus on the breaking of ergodicity as
the key to discrete time crystals and the delaying of ergodicity as the
source of numerous  phenomena that share many of the properties of discrete time crystals, including the AC Josephson effect, coupled  map lattices, and Faraday waves.
Theoretically, there exists a diverse array of strategies to stabilize time crystalline order in both closed and open systems, ranging from localization and prethermalization to dissipation and error correction. 
Experimentally, many-body quantum simulators provide a natural platform for investigating signatures of time crystalline order; recent work utilizing trapped ions, solid-state spin systems, and superconducting qubits will be reviewed.  Finally, this Colloquium concludes by describing outstanding challenges in the field and a vision for new directions on both the experimental and theoretical fronts.

\end{abstract}

\maketitle

\tableofcontents{}

\section{Introduction: Spontaneous Breaking of Time Translation Symmetry}
\label{intro}

Spontaneous symmetry breaking is a remarkable collective phenomenon:
an assembly of constituents, each interacting with only its nearby neighbors,
manages to align its behavior across large spatial and temporal separations.
The concept  has  wide-ranging applications, from  crystalline and magnetic ordering to superfluidity, superconductivity, and the generation of particle masses.
Recently there has been a burst of activity surrounding the spontaneous breaking of symmetries that involve time translation ~\cite{shapere2012classical,wilczek2012quantum,sacha2015modeling,khemani2016phase,else2016floquet,yao2017discrete}.
These explorations have exposed new phenomena and new opportunities, but also important subtleties. In this Colloquium, we will discuss both the resulting sharpening of theoretical concepts and the discovery of previously unsuspected new phases of matter.
We will also discuss open questions and potential applications.

Landau-Ginzburg theory is the starting point for many theoretical treatments of spontaneous symmetry breaking.
In a simple but representative example of this framework, one considers the theory of a complex scalar field $\phi(x, t)$ whose equations are invariant under a phase transformation $\phi \rightarrow e^{i\lambda} \phi$.  Physically, $\phi$ might represent the field associated with the creation and annihilation of bosonic particles (e.g., $^4{\rm He}$ atoms), $s$-wave spin singlet Cooper pairs, or the distribution of a planar spin density.
Within the Landau-Ginzburg paradigm, a key role is played by the energy (or more generally the free energy)  of a field configuration.
The simplest energy functional  consistent with  symmetries is given by:

\begin{equation}\label{ssbPotential}
 V(\phi)  =   b_4 | \nabla \phi| ^4 + b_2  |\nabla \phi| ^2
 + a_2 |\phi|^2 + a_4|\phi|^4 = H = -L ,
\end{equation}
where $\{a_2,a_4,b_2,b_4\}$ are real-valued parameters while $H, L$ denote the Hamiltonian and Lagrangian densities, respectively.   
Taking $a_4, b_4 > 0$  insures stability against rapid spatial variations and large fields.   

If $a_2 <0$, states with $\langle \phi \rangle \neq 0$ are energetically favorable, spontaneously breaking the phase rotation symmetry. If $b_2 < 0$, states with $\langle \nabla \phi \rangle \neq 0$ are favorable,  spontaneously breaking the spatial translation symmetry and indicating the emergence of spatial patterns. By reducing the  symmetry, the system gains a non-zero energy per unit volume.  
In the limit of infinite volume, the symmetry of all physically realizable states is less than the symmetry of the original equations.  
This is the mechanism of spontaneous symmetry breaking in equilibrium.

%

Physicists are accustomed, in many contexts, to treating time and space on an equal footing.   Thus, one could imagine generalizing Eq.\,(\ref{ssbPotential}) to include time derivatives.  In a Lagrangian framework, it is natural to consider adding the leading order kinetic terms,
\begin{equation}\label{tXtalL}
    L_{\rm kin.} =~c_4 |\partial_t \phi|^4 + c_2 |\partial_t \phi|^2,
\end{equation}
which corresponds to the energy density $H_{\rm kin.} = 3 c_4 |\partial_t \phi|^4 + c_2 |\partial_t \phi|^2$.
For $c_2 < 0, c_4 > 0$, it is energetically favorable to have $\langle \partial_t \phi \rangle \neq 0$,
implying the spontaneous breaking of time translation symmetry.

 This line of reasoning suggests that spontaneous time-translation symmetry-breaking (\TTSB{}) is straightforward to achieve.
 But subtleties abound~\cite{shapere2012classical, wilczek2012quantum}. Quantum mechanics requires Hamiltonians, and for non-singular Hamiltonians, Hamilton's equations, $\partial_t \phi = \partial_p H, \partial_t p = - \partial_\phi H$, imply that a system is stationary at any energy minima. Thus a ground state cannot exhibit periodic oscillations.
The behavior of  Eq.~\eqref{tXtalL} is able to evade this this conclusion  because $\partial_t \phi$ is not a single-valued function of the canonical momentum [i.e., $p(\partial_t \phi) = \frac{\delta L_{\rm kin.}}{\delta  \partial_t \phi}$
does not have a unique inverse $\partial_t \phi(p)$], and consequently, neither is the Hamiltonian.\cite{shapere2012classical}  This is not necessarily an insurmountable problem.  Indeed for such singular Hamiltonians it is often possible to construct consistent (in particular, unitary) quantum theories that realize the corresponding classical dynamics in the limit of large quantum numbers~\cite{shapere2012branched,shapere2012classical,zhao2013hamiltonian,chi2014single,choudhury2019branched}. Moreover, singular Hamiltonians that support time-dependent minima can arise as limiting cases of non-singular Hamiltonians as appropriate parameters are taken  large.  A limiting theory can be valuable, because it is often more tractable than the full theory, while being accurate within a  large range of parameter space ~\cite{shapere2019regularizations,dai2019truncated,alekseev2020provenance,dai2020classical}.  However, away from this limit several authors  \cite{bruno2013impossibility,nozieres2013time, watanabe2015absence} have argued that persistent oscillations cannot arise in a quantum system that is in equilibrium  with respect to a  local Hamiltonian.
 
 These subtleties suggest that the application of the
 Landau paradigm to \TTSB{}  may not be as easy as Eq.~\eqref{tXtalL} would lead us to believe.
 An additional necessary ingredient for \TTSB{} is
 a robust mechanism for \emph{ergodicity breaking}
 which enables the dynamics of the system to ``remember'' the initial condition (e.g.~the complex phase of $\phi$) out to infinite times.
Spontaneous symmetry breaking is a special case of ergodicity breaking. Focusing on this aspect of
spontaneous symmetry breaking leads to the following
formulations of \TTSB{}: it is a form of ergodicity breaking in which a generic ensemble of initial conditions exhibits persistent oscillations with a temporal phase shift that remembers the initial condition (cf.~\eqref{eq:def_TTSB}).

 \begin{figure*}
\centering
\includegraphics[width=1.0\linewidth]{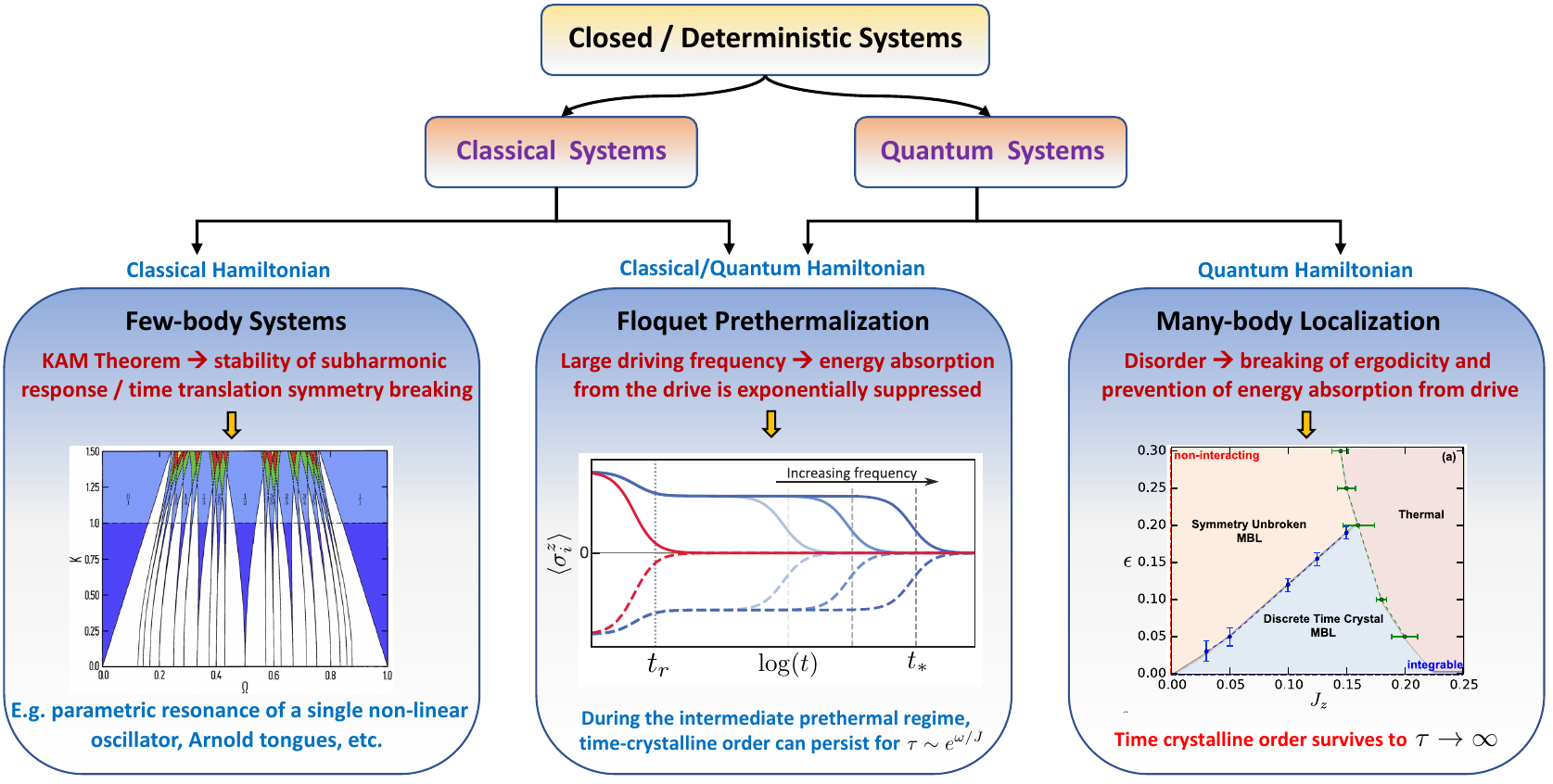}
\caption{Schematic depicting strategies for stabilizing time crystalline order in periodically-driven, closed  systems evolving via deterministic dynamics. In few-body classical systems, such as a parametrically driven non-linear oscillator, stable subharmonic responses are ubiquitous. 
This stability can be understood as a time-dependent extension of the Kolmogorov-Arnold-Moser (KAM) theorem~\cite{kolmogorov1954conservation,
moser1962invariant, Arnold2009},  which proves that quasiperiodic orbits of dynamical systems remain robust to  small perturbations (Sec.~\ref{sec:shaken_pendulum}).
There is no quantum analog to this classical few-body strategy. 
For many-body systems, both classical and quantum dynamics can exhibit Floquet prethermalization in the limit of large driving frequencies (Sec.~\ref{prethermal}).
During an intermediate prethermal window of time (i.e.~before Floquet heating occurs), the system can exhibit discrete time crystalline order.
The lifetime of this order scales exponentially with the frequency of the drive, $\tau \sim e^{\omega_D/J}$.
In strongly disordered, quantum many-body systems, a phenomenon known as many-body localization can occur. 
This prevents Floquet heating (see Sec.~\ref{ergodicity_breaking}) and also provides a mechanism for the many-body system to avoid becoming ergodic.
By breaking ergodicity, time crystalline order can persist to infinite times in the thermodynamic limit. 
Since many-body localization relies upon the discreteness of quantum mechanical levels, there is no classical analog to this strategy. 
} 
\label{fig:closed_system}
\end{figure*}

In recent years, a  non-equilibrium  route to \TTSB{} has been discovered in the context of Hamiltonians periodically driven at frequency $\omega_D$.
Such ``Floquet'' systems \cite{Floquet:1883} feature a reduced, discrete time-translation symmetry  $H(t + 2 \pi / \omega_D) = H(t)$, where $T_0 = 2 \pi / \omega_D$ is the driving period.
When a simple harmonic oscillator is driven at frequency $\omega_D$, it responds at frequency $\omega_D$, regardless of its natural frequency.
This is even true of many non-linear systems that are far more complex: when driven at frequency $\omega_D$, their observable properties respond at integer harmonics of $\omega_D$, regardless of their detailed structure.
In this case, the observable behavior evolves with the same discrete time-translation symmetry as the underlying equations of motion.
In contrast, a system is said to exhibit a \emph{subharmonic} response if there are  properties which oscillate at frequency $\omega_D / m$ for some integer $m>1$.
Such a subharmonic response  spontaneously breaks the discrete time-translation symmetry down to the smaller subgroup $t\rightarrow t+ 2\pi n/\omega_D$ with $n\in m\mathbb{Z}$.
If this \TTSB{} is stable to perturbations, the system is a ``discrete time crystal''~\cite{sacha2017time,guo2020condensed,sacha2020time,else2020discrete,khemani2019brief}.
Owing to the periodic drive, it is not obvious how to stabilize such behavior.
Within the Landau paradigm, spontaneous symmetry-breaking phases retain their order in the face of fluctuations (both thermal and quantum) by virtue of the energy penalty associated with misaligned regions.
But this is not possible for a discrete time crystal, since energy is not conserved in a driven system.
The oscillations in  distant parts of the system must remain in lockstep even though the energy penalty for failing to do so can easily be over-ridden by the energy supplied by the drive.

\begin{figure*}
\centering
\includegraphics[width=1.0\linewidth]{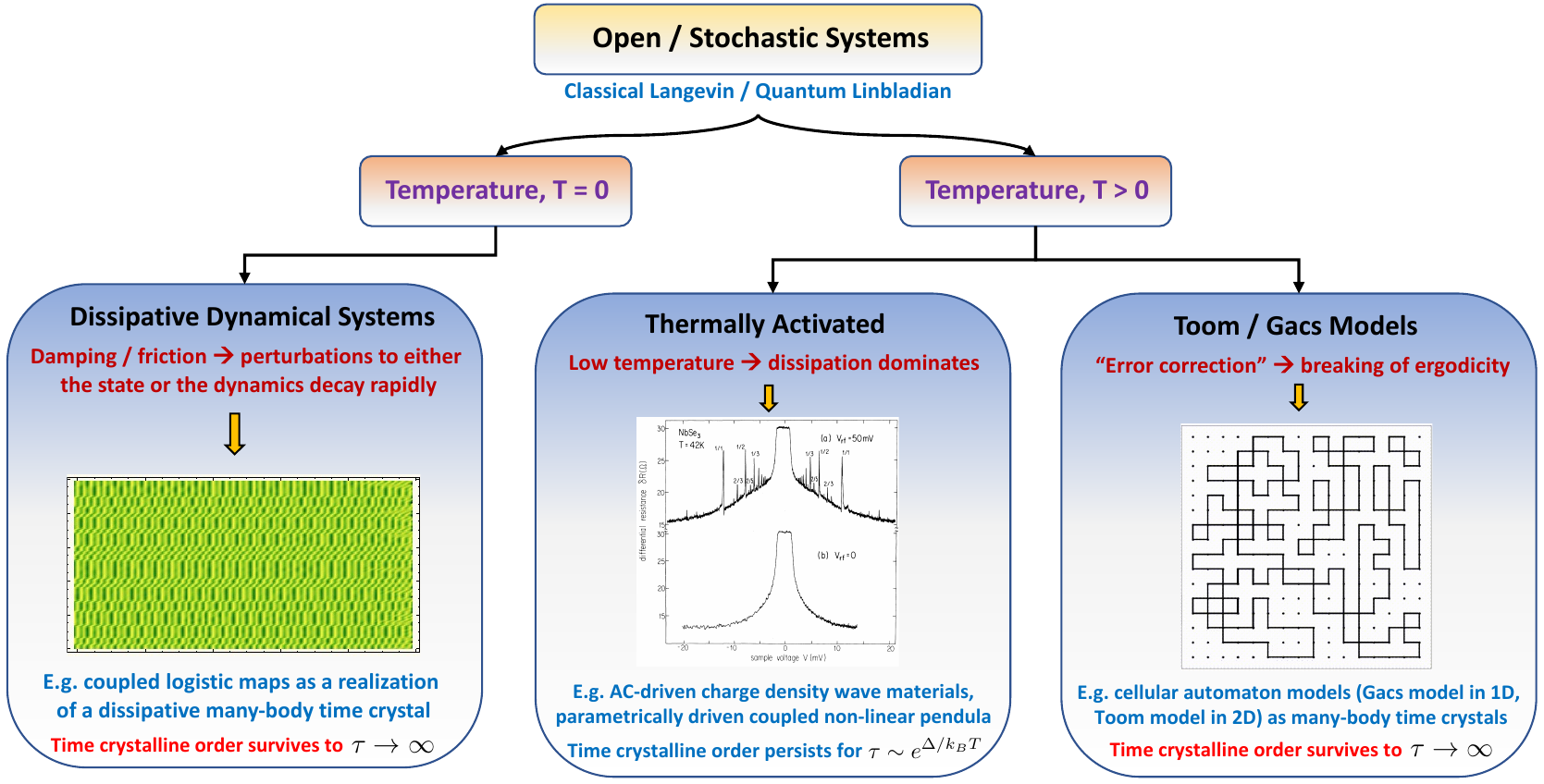}
\caption{Schematic depicting strategies for stabilizing time crystalline order in periodically-driven, open  systems evolving via stochastic dynamics. 
Dissipative non-linear dynamical systems have long been known to exhibit stable, many-body time translation symmetry breaking.
Indeed, coupled  map lattices can  satisfy all of the stated requirements for being a discrete time crystal. 
The key ingredient for the stability of \TTSB{}  in this setting is a generalized version of dissipation---effectively, one can think of the microscopic dynamics in such systems as being coupled to a zero temperature bath (alternatively, one can say that the microscopic dynamics are not information- or measure-preserving). 
This ensures that a finite volume of initial conditions contracts towards, for example, a period-doubled  fixed point (Sec.~\ref{sec:illustrative}). 
The stability of time crystalline order in many-body systems coupled to a finite temperature bath are significantly more subtle. For example, in classical Langevin dynamics or quantum Linbladian dynamics, at any $T>0$,   dissipation always comes with noise. 
In such systems (e.g.~finite temperature, parametrically-driven, coupled non-linear pendula), at low temperatures, time crystalline order can survive for an  ``activated'' time-scale $\sim e^{\Delta/k_B T}$ (Sec.~\ref{sec:activated_tc}). 
Somewhat remarkably, by considering more generic stochastic dynamics (i.e.~probabilistic cellular automata), it is possible to realize finite-temperature, time crystalline order with an infinite lifetime (Sec.~\ref{sec:toom}).
} 
\label{fig:open_system}
\end{figure*}

The aim of this Colloquium is to give a unifying perspective on various non-equilibrium mechanisms for discrete \TTSB{} in  closed (Fig.~\ref{fig:closed_system}) and open (Fig.~\ref{fig:open_system}) systems. 
By centering our discussion in the language of dynamical systems, we hope to highlight the connections --- and contrasts --- between the rich literature on subharmonic responses in dynamical systems and more recent developments in the context of closed quantum systems.
Until recently, many-body systems exhibiting \TTSB{},
such as period doubling in coupled  map lattices, were only found in open systems that relied fundamentally on dissipation.
They could be viewed as a type of engine: the energy supplied by the frequency-$\omega_D$ drive is converted to frequency $\omega_D / m$ motion while releasing heat to a cold bath.
With the discovery of many-body localized (MBL) discrete time crystals, a qualitatively different form of
\TTSB{} was discovered: one that does not generate any entropy at all.

In an MBL time crystal, the time scale  of \TTSB{} oscillations diverges exponentially as the system size alone is increased.
However, for practical purposes it is also of interest to know if there are other physical parameters whose limiting behavior can drive such an exponential increase, for example,  the drive frequency $\omega_D \to \infty$; the temperature   $T \to 0$; or the particle density $n \to \infty$. 
This will lead us to the notion of prethermal, activated, and driven-BEC time crystals, respectively. 
A key tool which is common to both of these ``exponentially good'' time-crystals and to their  MBL counterpart is the emergence of an effective time-independent \emph{Floquet Hamiltonian}, $H_{\textrm{eff}}$, which governs the dynamics out to exponentially long time scales.
Crucially, $H_{\textrm{eff}}$ can exhibit emergent symmetries that are protected by the underlying time-translation symmetry of the drive. 
In a pleasing return to form, \TTSB{} can then be understood in terms of the breaking of these emergent symmetries and as an application of the Landau paradigm not to $H(t)$, but  rather to $H_{\textrm{eff}}$.

In the setting of Floquet dynamics,  the environment couples to the system only via a coherent drive.
But in  any physical experiment, the environment itself is a many-body system, and thus, coupling to it inevitably comes along with dissipation and noise.
Indeed, all experiments on discrete time crystals to date show their fingerprints.
The possibility of perfect \TTSB{} in the presence of a drive, dissipation, \emph{and} noise  remains an open question.

The outlines of a possible answer to this question are provided by deep results from theoretical computer science and non-equilibrium statistical physics. 
As emphasized, \TTSB{} is fundamentally a form of ergodicity breaking, and there is a  history of rigorous results from these communities which show that ergodicity breaking can be generically stable to noise~\cite{toom1980, gacs2001reliable}.
In that context, the motivation was to understand whether ``reliable systems can emerge from unreliable components''~\cite{von2016probabilistic}.
The answer to this question is inextricably linked to the physical possibility of the most radical form of ergodicity breaking of all: classical and quantum error correction.
From this perspective, if purely dissipative time crystals are a form of engine, and MBL time crystals a type of idealized perpetual motion, then \TTSB{} in open systems is an embryonic example of an error-corrected  computer program: repeatedly applying the same ``$\textrm{NOT}$'' operation to all registers, the computer settles into the period-2 output $... 0101010101...$.

\section{What defines a time-crystalline phase of matter?}
\label{sec:what_defines}

 Discrete \TTSB{} leads to regular and long-lived oscillations with a period which is a multiple of the drive's.
 Oscillations in general, however, are ubiquitous throughout nature, so in this section we will formalize when such behavior signals the emergence of a genuine \emph{phase} of matter.
 Our starting point is any dynamical system \cite{birkhoff1927dynamical, katok1997introduction, strogatz2018nonlinear} whose state $x$  evolves under a discrete-time update rule $\Phi: x \to \Phi(x)$.
The state of the system at time-step $t$ is thus $x(t) = \Phi^{(t)}(x(0))$, where the superscript denotes iteration. 
The time-independence of $\Phi$ implies there is a discrete time-translation symmetry. 
While the notation highlights a discrete time-step, $\Phi$ may arise from viewing a continuous-time system ``stroboscopically.''
In the context of classical and quantum dynamics defined by time-periodic Hamiltonian $H(t + T) = H(t)$, $x$ is the classical or quantum state, and $\Phi$ corresponds to integration of Hamilton's equations or the Schrodinger equation over one period of the drive.

The dynamics exhibit $m$-fold time-translation symmetry breaking if there exists a local observable $O$ which exhibits periodic oscillations out to infinite times for a measurable volume of initial conditions $x$:
\begin{align}
 \lim_{\tau \to \infty} \frac{1}{\tau} \sum_{n=1}^\tau O(\Phi^{(m n)}(x)) \neq  \lim_{\tau \to \infty} \frac{1}{\tau} \sum_{n=1}^\tau O(\Phi^{(m n + p)}(x))
 \label{eq:def_TTSB}
\end{align}
where $0 < p < m$ corresponds to the $m$-phases of the orbit,  with equality restored for $p=m$.
A time-crystal thus ``remembers'' which of $m$ initial conditions it is in.
This implies that \TTSB{} is a particular form of \emph{ergodicity breaking}~\cite{sinai1959notion, cornfeld2012ergodic, walters2000introduction}: the time-averaged behavior of the $m$-fold iterated map $\Phi^{(m)}$ depends on the initial condition.
This behavior has been called ``asymptotic periodicity'' in the literature on classical many-body dynamical systems \cite{lasota1984asymptotic, losson1996thermodynamic, lasota2013chaos}.
While Eq.\eqref{eq:def_TTSB} implies infinitely long-lived oscillations, we may also consider a relaxed condition in which the time scale $\tau$ remains finite but diverges exponentially with a control parameter such as the drive frequency or  temperature. This will be the case for ``prethermal'' (Sec.~\ref{prethermal}) and ``activated'' (Sec.~\ref{sec:activated_tc})  time-crystals, respectively.

We emphasize that \TTSB{} does not require that each state $x$  itself undergoes perfectly periodic motion, only that there is an observable $O$ which oscillates on average.
For example, in the context of a single variable $x$, the observable $O(x) = \textrm{sign}(x)$ might exhibit  regular oscillations even while the motion of $x$ itself is quite chaotic.  
This  accounts for the fact that we rarely have direct access to the microstate of the system, only coarse-grained measurements.  
This limitation also motivates the requirement that oscillations be observable for a finite volume of initial conditions, over which we rarely have exact  control.
Taken together, an equivalent formulation of Eq.\eqref{eq:def_TTSB} takes a statistical point of view by considering the evolution of  \emph{distributions} over microstates $\rho(x)$; we  return to this formulation in our discussion of stochastic systems in Sec.~\ref{sec:open}.

Even a single degree of freedom can trivially exhibit \TTSB{}: take for example the map $\Phi(x) =  -x$.
The problem  gains its richness when we demand \emph{stability}:  \TTSB{} is a property of a dynamical \emph{phase} of matter if it is  robust to any small locality-preserving perturbation of either the initial condition $x$ or the dynamics $\Phi$.
Under the perturbation $\Phi(x) \to  -x(1 - \epsilon)$, for example,  oscillations are damped at  times beyond $\tau \sim \epsilon^{-1}$, and \TTSB{} is destroyed.

When discussing stability, we need to be clear which class of perturbations we demand stability against. 
Correspondingly this defines different possible \emph{classes} of \TTSB{} phases --- Hamiltonian, unitary, Langevin, quantum Lindladian, etc. --- which would exhibit \TTSB{} robust to arbitrary small perturbations within that dynamical class. 
Depending on how broadly or narrowly we define stability, we will have more or fewer examples, but weaker or stronger implications.  

\subsection{Illustrative examples}
\label{sec:illustrative}
The requirements surrounding Eq.~\eqref{eq:def_TTSB} contain several subtleties, so it will prove helpful to walk through several illustrative examples of ``is X a time-crystal'' with these in mind.
We first focus on  examples which do exhibit  oscillations (and are interesting in their own right), but which do not satisfy the strict requirements of Eq.\eqref{eq:def_TTSB}.

Consider first the dynamics of an undriven oscillator  $H = \frac{p^2}{2m} + \frac{\omega_0^2}{2 m }q^2 + \frac{\epsilon}{4} q^4$.
Obviously any initial condition $x = (q, p)$ will oscillate forever - do such oscillations constitute a discrete time-crystal?
Since $H$ is time-independent, for the purposes of testing Eq.~\eqref{eq:def_TTSB} we may choose to stroboscopically observe the dynamics at whichever period $T$ we suspect harbors the oscillations (say $T = 2 \pi / \omega_0$).
However, stability requires us to account for the non-linearity $\epsilon$, which causes the oscillation frequency to depend on the amplitude of the initial condition. 
As a result, there is no fixed period $T$ for which Eq.~\eqref{eq:def_TTSB} is satisfied for a \emph{finite volume} of initial conditions: a generic ensemble of states will instead dephase, with no \TTSB{} after some characteristic time $\tau$.

The time scale $\tau$ for such dephasing may be large, in which case the dynamics may \emph{appear} to exhibit \TTSB{} in practice even while they do not satisfy Eq.~\eqref{eq:def_TTSB} in principle. 
Consider for example the AC-Josephson effect, 
\begin{align}
H = -E_J \cos \phi + 2 e V n + \frac{n^2}{2C}
\label{eq:JJ}
\end{align}
Here $\phi$ is the superconducting phase difference across a Josephson junction, which is conjugate to the Cooper pair number $n$; $E_J$ is the Josephson energy for the tunneling of a pair across the junction, $C$ is the junction capacitance, and $V$ is the voltage difference across the junction.
For a large junction the capacitance approaches $C = \infty$, and the equations of motion then give $\phi(t) = \phi_0 + 2 e V t$, implying the supercurrent  $I_s = \dot{n} = E_J \sin(\phi)$ oscillates indefinitely in response to a DC voltage.
However, for the generic case in which $C$ is finite, a shift of variables from $n \to n - 2 e V C$ brings $H$ to the form of a pendulum whose  behavior is similar to the non-linear oscillator discussed above.
The time scale for the resulting dephasing $\tau = \sqrt{C / E_J}$ may be large - though it is worth pointing out that this dependence is not an exponential.
In experiments, Josephson junctions are generally resistively shunted, and  related  interesting phenomena arise in this open context as will be discussed in Sec.~\ref{sec:open}.

Another interesting limit of Eq.\eqref{eq:JJ} is $E_J = 0$, in which case $H$ has an internal $U(1)$ phase symmetry which guarantees that $\dot{\phi} = 2 e V + n / C$ remains constant.
In this case $T = 2 \pi / \dot{\phi}$
This is an example of \TTSB{} which ``piggy-backs’’ on the  spontaneous symmetry breaking of an unrelated internal symmetry.
Equivalent examples are superfluids at non-zero chemical potential or an $XY$ magnet in a perpendicular field.
Such systems have a $U(1)$ order parameter that precesses in time for generic initial conditions.
 The precession is unobservable (for example $I_s = 0$) unless the $U(1)$ symmetry, which makes all angles equivalent, is broken. 
But in the absence of $U(1)$ symmetry, there is no barrier to energy dissipating from the macroscopic precession of the order parameter into the internal motion of its many-body constituents.
At long times the system then relaxes to a state (e.g., vanishing particle number in the superfluid or spin aligned with the field in the magnet) wherein macroscopic motion ceases.  To navigate this dichotomy, one must shield the system from explicit symmetry breaking apart from brief, intermittent measurement events, as in the experiment of \onlinecite{Urbina82}.
In this case, \TTSB{} is stable only to perturbations which preserve the internal symmetry. 
In contrast, we will find that a stronger form of \TTSB{} is possible which depends  only on the time-translation symmetry itself. 

A  many-body example which does satisfy all the requirements surrounding Eq.~\eqref{eq:def_TTSB} is period doubling in deterministic \emph{dissipative} systems.
An iterated map $x\rightarrow  f(x)$, such as the logistic map,\cite{strogatz2018nonlinear} provides an idealized model of the unit-time evolution of such a system. Taking for example $f(x)=-x(1+a-{x^2})$,
there is a basin of attraction of initial states that settle into a limiting oscillation between $x=\pm\sqrt{a}$ when $a>0$. This simple one-body example can be promoted to a many-body model (a ``coupled map lattice'') by considering an array of variables $x_i$ which evolve under $\Phi: x_i \to f(x_i) + v(x_{i-1}, x_i, x_{i+1})$ with some generic local interaction $v$.
The behavior of coupled map lattices  is very rich, since they include cellular automata \cite{gutowitz1991cellular} as a special case~\cite{kaneko1984period,kapral1985pattern,bunimovich1988spacetime,kaneko1992overview,kaneko1987transition}. Coupled map lattices can support collective subharmonic responses \cite{losson1995phase,losson1996thermodynamic, gielis2000coupled}
which are stable to smooth perturbations of $f$ or $v$.

\section{Floquet Hamiltonian systems}

As noted above, if we consider  systems with dissipation but no noise, as for example a coupled  map lattice, it is relatively straightforward to obtain stable, many-body \TTSB{}. This is the classical zero-temperature limit
of the general open system problem that we discuss at length in Section \ref{sec:open}. These systems illustrate that the existence and stability of time crystals depends strongly on the class of dynamical system considered (open, closed; classical, quantum).
The presence of dissipation leads to a fixed-point orbit about which the linearized dynamics have eigenvalues $|\lambda| < 1$, so that a finite volume of initial conditions contracts towards the fixed point.
In contrast, for the canonical transformation which arises from integrating Hamilton's equations over one Floquet period, the eigenvalues  always come in conjugate pairs $\lambda_1 \lambda_2 = 1$ in order to ensure that the phase-space volume $dq \wedge dp$ is preserved in accord with Liouville’s theorem. In the section, we discuss why stable \TTSB{}, i.e. an infinitely-long-lived time crystal, is not expected to occur in such a situation.

\subsection{Faraday waves} 
\label{sec:faraday_waves}

To illustrate why Hamiltonian structure makes the possibility of \TTSB{} considerably more challenging, it is helpful to consider the concrete example of Faraday-wave instabilities in shaken surface waves. Faraday observed that when a container of water with a liquid-air interface is shaken vertically at frequency $\omega_D$, surface-waves develop which oscillate at frequencies $\omega = \frac{n}{m} \omega_D$, which are rational subharmonics of the drive~\cite{faraday1831xvii, rayleigh1883vii}.  The sub-harmonics can be understood as standing-wave modes that are forced via \emph{parametric resonance}~\cite{rayleigh1883xxxiii}.
    A precise mathematical understanding of the instability is obtained by linearizing the incompressible Euler equations for surface waves \cite{benjamin1954stability} in the presence of periodic acceleration. This reduces the problem to a set of ``Mathieu equations'' for each Fourier component of the surface height $q$, 
    \begin{align}
    \ddot{q}_k = -[\omega_k^2 + \delta_k  \cos(\omega_D t)] q_k    
    \label{eq:matthieu}
    \end{align}
    where $q_k$ is the amplitude of the surface-wave at wave-vector $k$, $\omega_k$ is the natural frequency of the surface wave, and $\delta_k$ is proportional to the driving amplitude.
The Mathieu equation features an exponentially-growing solution (which precisely looks like a subharmonic response) of the approximate form $\sim e^{t \Gamma} \cos(t \omega_D /2 + \theta )$ when a mode satisfies $\omega_D \sim 2 \omega_k$~\cite{mclachlan1947mathieu}.
    As we will discuss, the non-linearity then regulates the blow-up and stabilizes \TTSB{}.
    Since the Euler equations are Hamiltonian \cite{olver1982nonlinear}, it might seem that Faraday-wave instabilities then provide a Hamiltonian example of \TTSB{}.
    
 However, while this analysis implies a \emph{linear} subharmonic instability, it does not resolve whether the motion at long times is an example of \TTSB{} in the strong sense.
 While the  non-linearities which are generically present have the favorable property of regulating the exponential blow-up, they also couple different $k$-modes.
 A generic initial condition will contain some energy in these high-$k$ modes, which will then act as an effectively noisy force on the motion of the $k$ mode in which one is hoping to observe stable  \TTSB{}. 
 The key question can then be summarized as follows: Does \TTSB{} survive when treating the Faraday wave problem as a genuine non-linear many-body system, and if not, what governs the autocorrelation time of the long-lived subharmonic response  manifestly seen in experiments?

\subsection{Floquet Hamiltonians and emergent symmetries \label{sec:floq_H_emergent}}
In order to answer this question, it will prove helpful to walk through a specific example and to develop some intuition along the way. 
To this end, we will work through the example of a shaken non-linear pendulum (i.e.~a parametrically driven non-linear oscillator); for a single pendulum (and even in the few-body case), we will see that one naturally gets stable \TTSB{} as a consequence of the Kolmogorov-Arnold-Moser (KAM) theorem. \cite{kolmogorov1954conservation,
moser1962invariant, Arnold2009} 
In a truly many-body system, however, \TTSB{} becomes unstable.   We will argue that there is an intuitive,  general, mechanism underlying  this obstruction: ergodicity. 
Along the way, we will introduce and develop a particularly important construct --- namely, the  Floquet Hamiltonian --- which will provide a unifying framework for understanding many versions of \TTSB{} in closed Hamiltonian systems (both quantum and classical).

\begin{figure}
\centering
\includegraphics[width=1.0\columnwidth]{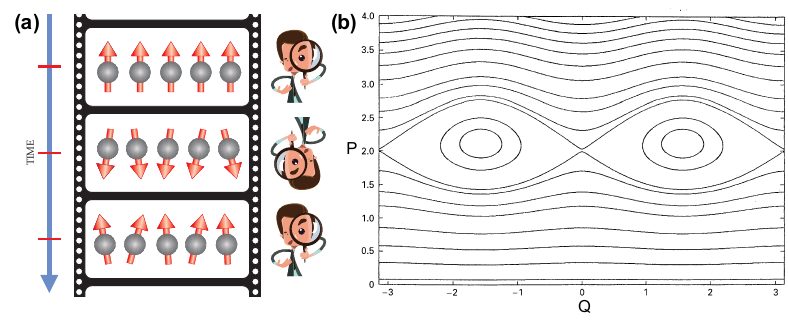}
\caption{(a) Schematic depiction of a time-dependent rotating frame, which transforms the original time-dependent $H(t)$ into the time-independent Floquet Hamiltonian, $H_F$. 
Time-translation symmetry breaking (here with period $m=2$) can be understood as  spontaneous symmetry breaking of an \emph{internal} $\mathbb{Z}_m$-symmetry of the Floquet Hamiltonian. Figure adapted from \cite{yao2018physicstoday}. (b) Contours of equal  ``quasi-energy'' $H_F(Q, P)$ for a  non-linear parametric oscillator  near a 2:1 resonance. The original coordinates of the oscillator $q, p$ are related to the coordinates $Q, P$ through a time-dependent canonical transformation given approximately by Eq.~\eqref{eq:K_approx}.  The two minima of $H_F$, at which $(Q, P)$ are time-independent, map back to the two exactly-period doubled orbits, $q_\ast(t) = - q_\ast(t + T_0)$. As predicted by Eq.\eqref{eq:zm_symmetry_floquet},  time-translation symmetry manifests as a $\mathbb{Z}_2$ symmetry $H_F(Q + \pi, P) = H_F(Q, P)$.
Figure adapted from Fig.~4 of \cite{zounes2002}.
} 
\label{fig:rotating_frame}
\end{figure}

Consider the Hamiltonian for a shaken non-linear pendulum:
\begin{align}
H(q, p, t) &= \frac{p^2}{2} +  \frac{\omega_0^2}{2}(1 + \delta \cos(\omega_D t) ) q^2 + \frac{\epsilon}{4} q^4.
\label{eq:H_pendulum}
\end{align}
To connect to the Faraday wave discussion above, the reader may think of this pendulum as a specific $k$-mode, where we are now explicitly accounting for the non-linearity but neglecting the inter-mode coupling.
The linearized equations of motion give rise to the Mathieu equation [c.f.~Eq.~\eqref{eq:matthieu}] and the goal, with respect to time crystalline order, would be to  prove that the non-linearity stabilizes \TTSB{}.

\subsubsection{The rotating frame: constructing the Floquet Hamiltonian}
\label{sec:rotating}

Near a period doubled solution of Eq.~\ref{eq:H_pendulum}, one expects on physical grounds that  trajectories should approximately take the form 
\begin{align}
q(t) - i p(t)  &=  \sqrt{2 P(t)} e^{i (\omega_D t / 2   + Q(t) ) }
\label{eq:K_approx}
\end{align}
where $Q(t), P(t)$ are \emph{slowly} varying in comparison to $\omega_D$.
The transformation that takes $(q, p) \to (Q, P)$ is a time-dependent canonical transformation, which we denote by $\mathcal{K}_0(t)$; this transformation can be thought of as taking the system to a particular ``rotating frame'' (Fig.~\ref{fig:rotating_frame}a).
Note that the period of the rotation is sub-harmonic,
$\mathcal{K}_0(m T) = \mathcal{K}_0(0)$ (in the period-doubled case, $m=2$).
In the rotating frame, the Hamiltonian, $H(Q, P, t)$ is still time-dependent, but the natural frequencies of the system (i.e.~those which govern the slowly-varying motion of $Q(t), P(t)$) are now off-resonant with $\omega_D$. 
Thus, one can attempt to follow up with a second transformation in order to generate a completely time-\emph{independent} ``Floquet Hamiltonian’’: $H_{F}$~\cite{holthaus1995classical, buchleitner2002non, else2017prethermal,else2020long}.
This second transformation  takes the form of a so-called Magnus expansion and the composite transformation is given by $\mathcal{K}(t) =  \mathcal{K}_{\textrm{magnus}}(t) \circ \mathcal{K}_0(t)$, again with $\mathcal{K}(m T)  = \mathcal{K}(0) $.
In summary, the strategy is to find a canonical transformation $\mathcal{K}(t): (q, p) \to (Q, P)$ such that the transformed Hamiltonian $H_F(Q, P)$ is time-independent. 
This idea that $H(q,p,t)$ might actually be equivalent to an equilibrium system (i.e.~governed by $H_F$) in an appropriate time-dependent rotating frame is referred to as ``crypto-equilibrium'' in \cite{yao2018physicstoday,else2020discrete}.

We emphasize that a-priori, the Magnus expansion may not converge, in which case the pair of objects $\{\mathcal{K}(t), H_F\}$ need not exist ---  this will  be the precise obstruction to stable \TTSB{} in the many-body Faraday-wave case. 
Note that in the quantum setting (which we will return to in just a few paragraphs), one can, in principle, always define a Floquet Hamiltonian by taking the logarithm of the Floquet unitary; thus, in this case, one should replace the ``need not exist'' from the previous sentence with the fact that $H_F$ will be \emph{highly non-local}.

As a preview, we note that the distinction between a true discrete time crystal (Sec.~\ref{MBLDTC}) and the prethermal discrete time crystal (Sec.~\ref{prethermal}) can precisely be understood as \emph{whether} the Magnus expansion converges. %
In the former case, it does and one has a well defined Floquet Hamiltonian, $H_F$, while in the latter case, it \emph{almost} does, leading to:
\begin{align}
H(Q, P, t) = H_\textrm{eff}(Q,P) + V(t).
\label{eq:Heff}
\end{align}
Here, $H_\textrm{eff}$ is an effective Hamiltonian that captures the system's stroboscopic dynamics for exponentially long time-scales and $V(t)$ is the (exponentially small) residual time-dependence which cannot be ``rotated'' away by the Magnus expansion.

The concept of a ``rotating frame'' transformation in the quantum case is completely analogous. 
In particular, within the quantum setting, the dynamics are captured by a unitary time evolution operator:
\begin{equation}
\label{eq:floquetop}
U(t_1, t_0) = \mathcal{T} \exp\left(-i\int_{t_0}^{t_1} H(t) dt \right),
\end{equation}
where $\mathcal{T}$ represents  time-ordering.
In this case, $\mathcal{K}(t)$ is a unitary transformation chosen such that the time-evolution operator $U(t_1, t_0)$   decomposes as
\begin{align}
U(t_1, t_0) = \mathcal{K}^{-1}(t_1) e^{-i (t_1 - t_0) H_F} \mathcal{K}(t_0)
\label{eq:floquet_decomposition}
\end{align}
with $\mathcal{K}(m T) =\mathcal{K}(0)$ and $H_F$ suitably local. $\mathcal{K}(t)$ is  referred to as the ``micromotion,'' since it contains the components of the dynamics at harmonics of $\omega_D/m$, while the ``slow'' part of the motion resides in $H_F$~\cite{shirley1965solution,zel1973scattering,bukov2015universal,rahav2003time,eckardt2015high,buchleitner2002non}.
When restricting to the stroboscopic dynamics, the ``Floquet unitary'' which implements a single discrete time step is given by  $U_\text{F} = U(T, 0) = \mathcal{K}^{-1}(T) e^{-i T H_F} \mathcal{K}(0)$.
To ensure that the connection to the notation in Eq.~\ref{eq:def_TTSB} is clear, we note that Floquet unitary, $U_\text{F}$, then furnishes the dynamical map $\Phi$ which acts on the quantum state $x = \ket{\psi}$.

\subsubsection{Time translation symmetry breaking as internal symmetry breaking of the Floquet Hamiltonian}
\label{sec:internal}

If the micromotion is periodic, $\mathcal{K}(T) =  \mathcal{K}(0)$, there can be no \TTSB{}, since the stroboscopic dynamics  become canonically equivalent to the un-driven dynamics of $H_F$, and \TTSB{} is forbidden in equilibrium~\cite{bruno2013impossibility,watanabe2015absence}.
But in a time-crystal, the frame  rotates subharmonically (Fig.~\ref{fig:rotating_frame}a), i.e.~$\mathcal{K}(m T) =  \mathcal{K}(0)$, and the  transformation $X = \mathcal{K}(0)  \mathcal{K}^{-1}(T)$ is non-trivial. 
The  transformation $X$ then plays a crucial role since it  permutes through the $m$ cycles of the sub-harmonic response. 
In the specific case of the shaken non-linear pendulum [c.f.~Eq.~\eqref{eq:K_approx}], for example, $X: (Q, P) \to (Q + \pi, P)$. We will expand upon this discussion in the next subsection (Sec.~\ref{sec:shaken_pendulum}). 

Under rather general conditions, the Magnus expansion can be constructed  to ensure that $X = \mathcal{K}(n T)  \mathcal{K}^{-1}( n T + T )$ is independent of $n$ \cite{ else2017prethermal,else2020long,machado2020long}, which then ensures that $X^m = \mathbb{I}$.
Substituting Eq.~\eqref{eq:floquet_decomposition} into the time-translation symmetry $U(T, 0) = U(2 T, T)$, one finds that
\begin{align}
X^{-1} e^{- i T H_F} X = e^{- i T H_F}.
\label{eq:zm_symmetry_floquet}
\end{align}
In other words, $X$ is an internal $\mathbb{Z}_m$ symmetry of the Floquet Hamiltonian~\cite{holthaus1995classical, buchleitner2002non, von2016absolute,else2017prethermal}.
While the emergence of an internal symmetry may seem mysterious, it is really just a re-expression of the original discrete time-translation symmetry in a frame which is rotated by $X$ at each step (Fig.~\ref{fig:rotating_frame}a). In \cite{else2020long}, this intertwining is referred to as ``twisted time-translation''. 
However, the emergence of such a symmetry was pointed out well before the recent interest in \TTSB; see for example  the extensive literature on non-spreading wavepackets in periodically driven systems.
    \cite{holthaus1994subharmonic, holthaus1995classical, buchleitner2002non}.

The existence of a Floquet Hamiltonian $H_F$ and an emergent symmetry $X$ immediately leads to the possibility of \TTSB{}.
In particular, suppose that the dynamics of $H_F$ spontaneously break the $X$ symmetry.
Intuitively, this means that the state space breaks up into $m$ ``superselection'' sectors which are permuted by the action of $X$ but which
are not connected by evolution under $H_F$. 
In the rotating frame, a trajectory thus gets stuck within a sector, and mapping back to the lab frame using $\mathcal{K}^{-1}(t)$, one finds that the average behavior will oscillate with period $m$.
More rigorously, symmetry breaking implies that there are $m$ different initial conditions $y_j = (Q_j, P_j)$ which are related by the symmetry, $y_{j+1} = X y_j$.
The time-averaged behaviors of these different initial conditions are distinct (in the rotating frame), e.g., there are local observables $O$ with $\bar{O}(y_j) \equiv \frac{1}{\tau} \sum^\tau_{n = 0} O(y_j(n T )) \neq \bar{O}(y_{j+1})$.
Transforming back to the lab frame, $x_j(t)  = \mathcal{K}^{-1}(t) y_j(t)$ and substituting in the definitions, one immediately finds that the definition of time-crystalline order [c.f.~Eq.~\eqref{eq:def_TTSB}] is satisfied. 

One thus arrives at a correspondence between $m$-fold \TTSB{} and the more familiar notion of  $\mathbb{Z}_m$-spontaneous symmetry breaking with respect to $H_F$ (in the rotating frame). 
This illustrates a general principle of all closed  Hamiltonian many-body time-crystals discovered thus far (Fig.~\ref{fig:closed_system}): In a Hamiltonian time-crystal, the discrete time-translation symmetry of $H(t)$ manifests as a spontaneously broken internal $\mathbb{Z}_m$-symmetry of the Floquet Hamiltonian $H_F$~\cite{else2017prethermal,von2016absolute,else2020long}.
In a sense this bring us back full circle to our starting inspiration, the Landau paradigm (Sec.~\ref{intro}), with the crucial replacement $H \to H_F$.

\subsubsection{The spectral perspective}
\label{sec:spectral_perspective}
While the above formalism applies equally well in the classical and quantum settings, in the quantum case an equivalent definition can be given in terms of the spectral properties  of the Floquet operator $U_F$ alone~\cite{holthaus1995classical, buchleitner2002non, sacha2015modeling}: a discrete time crystal is a phase of matter in which the  eigenstates of $U_F$ are \emph{necessarily} ``cat states,'' i.e.~superpositions of macroscopically-distinct states~\cite{else2016floquet}.
This is a direct generalization of the statement that when a conventional symmetry is spontaneously broken in an equilibrium system, the low-energy eigenstates that transform simply under the symmetry (e.g. eigenstates of the center or Cartan sub-algebra \cite{fulton2013representation} of the symmetry group) are necessarily cat states.

To unpack this definition we investigate the implications of Sec.~\ref{sec:internal} for the ``Floquet eigenstates''   $U_F |\varepsilon_F\rangle = e^{-i T \varepsilon_F }  |\varepsilon_F\rangle$.
Here the ``quasienergies'' $\varepsilon_F$ are defined modulo the driving frequency $\varepsilon_F \equiv \varepsilon_F + \omega_\textrm{D}$. 
As discussed, $U_F$ exhibits \TTSB{} when there is a decomposition $U_F = X e^{-i T H_F}$ for which $H_F$ spontaneously breaks the symmetry $[X, H_F] = 0$.
In the thermodynamic limit, spontaneous symmetry breaking  implies that the eigenstates of $H_F$ come in degenerate pairs permuted by the symmetry, $X \ket{\uparrow} = \ket{\downarrow}$ (we focus  on $X^2 = 1$ for simplicity) \footnote{In conventional symmetry breaking, only eigenstates  with energy densities below the symmetry-breaking temperature $T_c$ have this structure. However, if many-body localization is able to exist in Floquet Hamiltonians~\cite{ponte2015many,vsuntajs2020quantum,sels2021dynamical}, it is believed that symmetry breaking must manifest at \emph{all} energy densities~\cite{de2016absence,moudgalya2020perturbative,sahay2021emergent}.}.
However, in a finite system, the  eigenstates of $H_F$ must simultaneously diagonalize $X$, so they come in pairs $\ket{\pm} = \frac{1}{\sqrt{2}}( \ket{\uparrow} \pm \ket{\downarrow}$ with eigenvalues $\epsilon^{\pm} = \epsilon \pm \Delta$ which are split by an amount $\Delta \sim e^{-L/\xi}$ which is exponentially small in the system size $L$ if the system is to exhbit true \TTSB{}.
The eigenstates of $H_F$ are thus ``cat states,'' since the constituent $\ket{\uparrow / \downarrow}$ are macroscopically distinct.
The quasienergies of $U_F$ are then obtained by including their eigenvalue under $X$: 
$\epsilon^{+}_F = \epsilon +\Delta$, and  $\epsilon^{-}_F = \epsilon -\Delta + \omega_D/2$.
The spectrum of $U_F$ thus consists of pairs of cat states with eigenvalues that differ by $\omega_D/2$ up to exponential accuracy in $L$.

To relate this spectral property back to  Eq.~\eqref{eq:def_TTSB}, note that an initial state $\ket{x}$ will generically have amplitude on both the $\ket{+}$ and $\ket{-}$ states of each pair. 
Due to the eigenvalue pairing, the system will thus coherently oscillate at frequency $\omega_D/2$, up to a dephasing time $\tau \sim \Delta^{-1}$ which diverges exponentially in $L$.

    \subsubsection{Application to many-body parametric resonance}
    \label{sec:shaken_pendulum}
    
   Let us now see how the formalism from Secs.~\ref{sec:rotating},~\ref{sec:internal} plays out for a single parametrically driven non-linear pendulum [c.f.~Eq.~\ref{eq:H_pendulum}]. \onlinecite{chirikov1979universal, zounes2002} showed that $H_F(Q,P)$ indeed exists within a finite volume of phase space (calling it the ``resonance Kamiltonian''), and that courtesy of the KAM theorem, its existence is stable to small  perturbations. 
The contours of equal ``quasi-energy'' $H_F(Q, P)$ are shown in Fig.~\ref{fig:rotating_frame}b.
As aforementioned, we see that $H_F$ exhibits a $\mathbb{Z}_2$-symmetry, $X: Q \to Q + \pi$, which exchanges two local minima enclosed by a separatrix.
The detailed construction of $\mathcal{K}(t)$ is quite involved, but the \emph{approximate} intuition is as follows: One starts with the  canonical transformation in Eq.~\eqref{eq:K_approx}, and removes the residual time dependence  order-by-order through a sequence of  transformations whose convergence is guaranteed by the KAM theorem.
The two minima of $H_F$,   $Q, P = (\pm \frac{\pi}{2}, P_\ast)$, are fixed points of $H_F$; mapping back to the lab frame, one obtains two precisely period doubled orbits related by time translation, $\mathcal{K}^{-1}(t): (\pm  \frac{\pi}{2}, P_\ast)  \to q_\ast(t \pm T/2), p_\ast(t)$. Small deviations from the minima circulate around quasi-energy contours, corresponding to slow oscillations about the period doubled motion. 
Similar behavior is seen in other periodically driven problems, such as the kicked rotor model, whose stroboscopic motion $\Phi$ reduces to Chirikov's standard map,\cite{chirikov1979universal}
or a particle bouncing off an oscillating mirror\cite{holthaus1994subharmonic, holthaus1995classical}.    
    In this 0D case, the \TTSB{} arises because KAM stability allows  the quasi-energy  to ``split up'' the $\{Q,P\}$-space into disconnected basins (Fig.~\ref{fig:rotating_frame}b), rather than because of a collective phenomena.
    Furthermore in 0D there is long-range order in time, but not in space,  and as such the reader may prefer not consider this case a time crystal. 
    This analysis also  illustrates why the quantum version of a parametrically-driven non-linear oscillator will not feature infinitely long-lived \TTSB{}: while a quantum Floquet Hamiltonian with a $\mathbb{Z}_2$ symmetry may exist, for any finite barrier height there will be quantum tunneling between the two minima at a rate $\sim 1 / \tau$. This leads to a unique $\mathbb{Z}_2$-symmetric steady state, and the subharmonic response will have a correlation time of $\tau$.
    \cite{holthaus1994subharmonic, holthaus1995classical, buchleitner2002non, sacha2015modeling} analyze several examples of such quantum subharmonic responses and provide estimates of the tunneling rate, including a particle bouncing off an oscillating mirror, a topic we will return to in Sec.~\ref{sec:driven_BEC}.

    The fly in the ointment comes when we attempt to construct $H_F$ in the  many-body setting.
    In particular, let us consider  an array of coupled pendula by adding a nearest-neighbor interaction to Eq.~\ref{eq:H_pendulum}, 
\begin{align}
H(q, p, t) &= \sum_i \left[ \frac{p_i^2}{2 } + \frac{ \omega_0^2}{2} (1 + \delta \cos(\omega_D t) ) q^2_i + \frac{\epsilon}{4} q_i^4 \right] \nonumber \\
& \quad - g \sum_{\langle i, j \rangle}(q_i - q_{j})^2.
\label{eq:H_FK}
\end{align}
Physically, the reader can think of this as treating the pendulum  as a \emph{macroscopic} object composed of atoms (i.e.~the Frenkel-Kontorova model \cite{kontorova1938theory}). It differs from the Faraday problem in the details of the dispersion and non-linearity, but is otherwise conceptually similar.

If the initial condition is uniform, $q_i, p_i = q, p$, the problem exactly reduces to Eq.~\eqref{eq:H_pendulum} which we now know exhibits stable period doubling. 
But stability for a \emph{finite-volume} of initial conditions requires one to consider initial conditions of the form, $q_i = q_\ast + \delta q_i$, in which the $\delta q_i$ are small but independent. 
Using the language we introduced near Eq.~\ref{eq:matthieu}, there is now weight in the higher-$k$ modes, $q_k$, which are all-to-all coupled through the  $\sum_i q_i^4$ non-linearity, so the center-of-mass mode (i.e.~$k=0$) cannot be examined in isolation.
While one can attempt to employ the same time-dependent canonical transformation $\mathcal{K}(t)$ site-by-site, for generic initial conditions the coupling $g (q_i - q_j)^2$ will \emph{not} be exactly time-independent after this transformation; thus, since the transformed Hamiltonian $H(Q_i, P_i, t)$ remains time-dependent we cannot appeal to the conservation of $H_F$ to ensure stability.
One can attempt to remedy this by adjusting the canonical transformation via a perturbative expansion in $g$, but there is a rather general reason to expect this expansion will fail to converge in the thermodynamic limit: the ubiquity of ergodicity.

\subsection{Ergodicity, destroyer of  time-crystals} 
\label{ergodicity_breaking}

As we have discussed, Hamiltonian systems are distinguished from more general dynamical systems like coupled  map lattices by the existence of a phase-space volume element, $d \mu = dq \wedge dp$, which is invariant under the dynamics, i.e., the dynamics are  ``measure preserving.'' An invertible measure preserving system is said to be \emph{ergodic} if the only finite-volume subset $A$ which is invariant under the discrete-time  update  rule $\Phi$ (i.e.~$\Phi(A) = A $) is the \emph{entire} phase space.
An equivalent statement is that  the orbit of any volume $\bigcup_m \Phi^{(m)}(A)$ eventually fills all of phase space~\cite{cornfeld2012ergodic}.
Full ergodicity is incompatible with the existence of a time-independent $H_F(Q, P)$ because the conservation of the quasi-energy $H_F$ partitions phase space into $\Phi$-invariant quasi-energy contours (though the dynamics may be ergodic \emph{within} each contour). 
A time-dependent many-body system with a bounded state space is \emph{generically}  expected to be ergodic in the thermodynamic limit.
\footnote{The ideas emanating from Boltzmann's ``ergodic hypothesis'' postulate that time-\emph{independent} many-body systems are generically ergodic within each conserved energy shell.\cite{von1991boltzmann} This hypothesis relates to the Floquet setting by converting the time-dependent Hamiltonian $H(t)$ into a time-independent $\mathcal{H} = H(t) + \pi_t$, where  $(t, \pi_t)$ are an additional canonical pair  living on the cylinder $t \sim t + T$. If $\mathcal{H}$ is ergodic on an energy shell, then the stroboscopic dynamics of $H(t)$ are fully ergodic.}

Or, in the unbounded case: the generic fate of a driven, many-body Hamiltonian system is to heat up.
In an ergodic system, Birkhoff's Ergodic Theorem \cite{birkhoff1931proof, cornfeld2012ergodic}  equates a temporal average of the sort defined in Eq.~\eqref{eq:def_TTSB} with averages over the entire phase-space.
Thus \TTSB{} necessarily requires $\Phi^{(m)}$ to be non-ergodic:  a time-crystal ``remembers'' its initial condition for an infinitely long time because it is encoded in the phase of the period-$m$ oscillations. 
\TTSB{} is thereby deeply connected with another profound phenomena: ergodicity breaking. 

   When the interactions are weak, or the driving is strong, the \emph{time-scale} required to explore all of phase space can be very large, so  in practice, \TTSB{} may persist out to very long times $\tau$. 
   This behavior is sometimes referred to as slow \emph{Floquet heating}.  While energy is not conserved, in systems described by an effective Floquet Hamiltonian [Eq.~\eqref{eq:Heff}], the quasienergy $\langle H_{\textrm{eff}} \rangle$  will change slowly due to the residual drive $V(t)$, eventually allowing the system to explore its full phase space.
     We will return to such anomalously long time scales when  we  discuss ``prethermal'' time-crystals in Sec.~\ref{prethermal}.

The existence of such a long time scale is certainly one ingredient in Faraday's observation of period-doubling in a system which is Hamiltonian.  
An additional role is played by viscosity, which converts the high-$k$ oscillations of the surface into heat which is eventually dissipated into the environment.
When viscosity alone is  added to the equations of motion, the system is no longer measure preserving and becomes analogous to the coupled map lattices where true \TTSB{} is possible.
However, this might seem to be a contradiction, because viscosity is after all a phenomenological treatment of microscopic degrees of freedom which are themselves Hamiltonian! The key is that at any finite temperature (energy density), a microscopic bath invariably  generates \emph{noise} in addition to viscosity, as required by the fluctuation-dissipation theorem.
Understanding the fate of \TTSB{} in the presence of dissipation \emph{and} noise is a particularly rich direction; we will return to the possibility of such ``open'' time-crystals in Sec.~\ref{sec:open}.
    
    To summarize, in dissipative dynamical systems like the coupled map lattices, stable, many-body \TTSB{} is  possible, and  time-crystals in this context have a long history.
    But their realization in  measure-preserving systems which describe the universe at the microscopic level is far more subtle.
While stable \TTSB{} is possible in few-body classical systems such as a parametric resonator (i.e.~the shaken pendulum example in Sec.~\ref{sec:shaken_pendulum}), their existence in many-body systems requires, at minimum, a generic mechanism for breaking  ergodicity.
    In closed classical systems, this is not thought to be possible except when the system is fine tuned.
    This is where the magic of quantum mechanics comes in,  through the remarkable physics of many-body localization (MBL). MBL allows for stable ergodicity breaking in the thermodynamic limit, and where there is ergodicity breaking, there will be time-crystals.

\begin{figure*}
\centering
\includegraphics[width=5.0in]{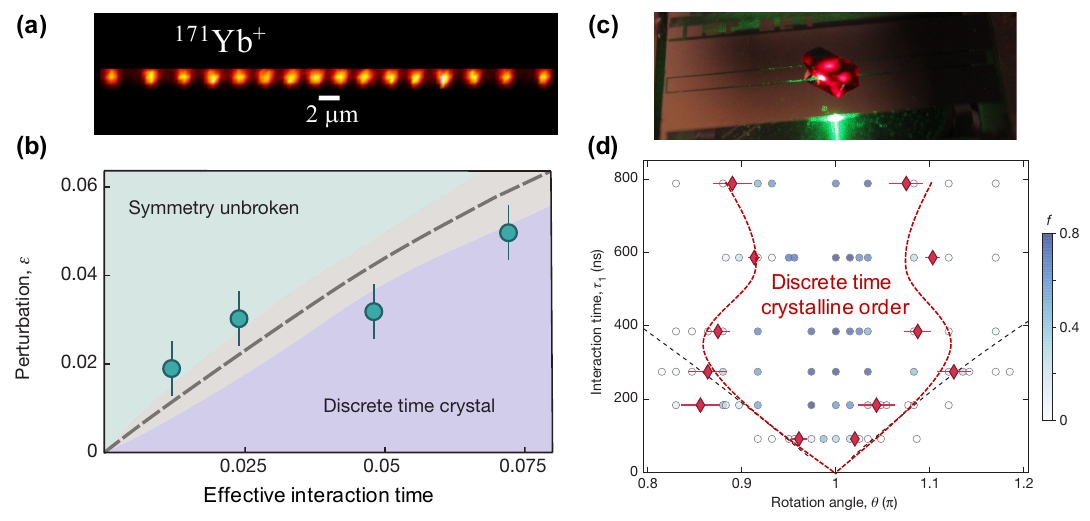}
\caption{(a) Signatures of time crystalline order were observed in a one dimensional chain of Ytterbium ions. In this system, the effective spin degree of freedom is encoded in two hyperfine ``clock'' states of a $^{171}$Yb$^+$ ion. (b) By varying the interaction strength (x-axis), which is effectively parameterized by an interaction time-scale, and the strength of the perturbation  on the $\pi$-pulse (y-axis), \cite{zhang2017observation} were able to observe the system's behavior change from a discrete time crystal phase to a time-translation symmetry unbroken phase. Figure adapted from Fig.~3(b) of \cite{zhang2017observation}. (c) \cite{choi2017observation} utilized  a diamond sample containing a high density of Nitrogen-Vacancy color centers ($\sim 15$~ppm)~\cite{choi2020robust}. In this system, the spin consists  of two $m_s$-sublevels and NV centers interact with one another via magnetic dipole-dipole interactions. 
The interplay between the long-ranged dipolar interaction and the three dimensional nature of the sample is conjectured to lead to critically slow thermalization~\cite{kucsko2018critical,ho2017critical}, during which discrete time crystalline order can be observed. 
(d) Experimentally measured phase boundary (red dashed line added as a guide) of the critical discrete time crystal as a function of the effective interaction strength and the $\pi$-pulse imperfection. Figure adapted from Fig.~3(b) of \cite{choi2017observation}.
} 
\label{fig:original_expts}
\end{figure*}

\section{Closed, periodically-driven quantum systems}
\label{DTCFloquet}

\subsection{Introduction to quantum Floquet phases}
\label{Floquet}

As noted above, ergodicity-breaking is necessary for the stability of a time crystal.
In fact, with the insights of the previous section in hand, one can utilize almost any generic,
robust form of ergodicity-breaking to stabilize a time crystal. However, this wasn't
immediately apparent when many-body localization \cite{nandkishore2015many,abanin2019colloquium} was originally discovered.

Many-body localization occurs in 1D and, possibly, 2D systems with strong quenched disorder and short-ranged interactions~\cite{smith2016many,choi2016exploring}. 
It represents a failure of the eigenstate thermalization hypothesis \cite{Deutsch91,Srednicki94} that occurs when disorder prevents equilibration by localizing the degrees of freedom of a system.
Locally-injected energy cannot spread throughout the system~\cite{nandkishore2015many,abanin2017recent,abanin2019colloquium}. MBL was
originally formulated as a property of highly-disordered
time-independent Hamiltonians $\hat{H}$ in which local observables fail to relax to their thermal average with respect to $e^{-\beta \hat{H}}$ and,
instead, the long-time behavior depends on the details of the initial state, a clear example of ergodicity breaking. 
It was subsequently realized that
MBL can also occur in 1D periodically-driven systems with strong disorder for appropriate driving parameters~\cite{ponte2015many,lazarides2015fate,abanin2016theory}.
In such Floquet-MBL systems, energy absorbed
during one part of the drive cycle must be returned before its completion~\cite{d2013many,ponte2015many,bordia2017periodically}. Consequently, the energy spreading and thermalization necessary for drive-induced Floquet heating cannot occur (Sec.~\ref{ergodicity_breaking}).

In the meantime, interest in Floquet systems had been re-energized by
seminal work demonstrating that periodic driving can induce topological properties in an otherwise nominally trivial system \cite{Inoue10,Lindner11a,WangY13,Jiang11b,Thakurathi13,vonKeyserlingk16a,vonKeyserlingk16b,potter2016classification,roy2016abelian,roy2017periodic,potirniche2017floquet}. In particular, \cite{Jiang11b} and \cite{Thakurathi13}  considered
periodically-driven versions of a Majorana chain in which the
time-dependent Hamiltonian toggles between
topological superconducting and trivial insulating states. These free fermion
models avoid Floquet heating by virtue of their integrability, which does not survive
generic perturbations involving interactions.
For such drives, the Floquet operator
takes on a particularly simple form, $U_\text{F} = \exp\!\Big[{i{c}\sum_j i\gamma^A_j \gamma^B_j}\Big]
\exp\!\Big[{i J\sum_j i\gamma^B_j \gamma^B_{j+1}}\Big]$~\cite{Prosen98,Thakurathi13}.
For $c$ near $\pi/2$, there are ``Floquet Majorana fermions'' at quasienergy $\pi$.
In other words, there are locally-indistinguishable states of even/odd fermion parity,
associated with edge modes. In an undriven system, Majorana edge modes would be degenerate (i.e. they would be zero modes), but the periodic driving 
splits their Floquet eigenvalues by $\pi$.

This property of the spectrum does not imply subharmonic response of the system for generic initial conditions (and, moreover, it is not stable to interactions). However, a Jordan-Wigner transformation
of the driven Majorana chain yields a driven Ising model 
that exhibits subharmonic oscillations of the Ising order parameter; moreover, by adding strong disorder leading to MBL, the system could be stabilized against interactions \cite{khemani2016phase}.  Initially it was not believed that the subharmonic response of the resulting
``$\pi$ spin glass'' could survive in the absence
of the Ising symmetry, implying that, in analogy to the discussions of the XY magnet in Sec.~\ref{sec:illustrative}, a time crystal stable against generic perturbations did not exist~\cite{khemani2016phase}.

This sentiment was overturned in \cite{else2016floquet}, which showed that discrete time crystals are a stable phase of matter in Floquet-MBL systems with only time translation symmetry.
In particular, a discrete time crystal 
in a closed system
must satisfy Eq.~\ref{eq:def_TTSB} in the absence of a bath
and in a manner that is stable against arbitrary weak perturbations
respecting time translation symmetry.
Other symmetries may or may
not be present, but Eq.~\ref{eq:def_TTSB} should hold even when
those symmetries are violated. 
These developments led rapidly to the first phase diagram for a time crystal and to proposals describing how to create and measure discrete time crystals in experiments~\cite{yao2017discrete}, culminating in the first observations of time-crystalline behavior~(Fig.~\ref{fig:original_expts})~\cite{zhang2017observation,choi2017observation}.

Thus far, the above discussions have focused on many-body localization as a possible strategy for stabilizing discrete time crystals. One of the caveats of using MBL is that the approach depends on the robustness of many-body localization itself. 
While MBL has purportedly been proven to exist in certain undriven, one-dimensional spin chain models with random local interactions~\cite{imbrie2016many}, its stability in $d>1$ and in long-range interacting systems remains an open question~\cite{yao2014many,de2017stability,luitz2017small,thiery2018many,khemani2021comment,yao2021reply}; to date, there is no mathematical proof for the stability of many-body localized \TTSB{} in a Floquet system in any dimension.
An alternate strategy for stabilizing DTC order, termed Floquet prethermalization, can occur in any dimension and in the presence of long-range interactions, so long as the driving frequency, $\omega_D$, is  sufficiently high.
To this end, in the remainder of this section focused on
closed systems (Fig.~\ref{fig:closed_system}), we will focus in subsection \ref{MBLDTC} on MBL discrete
time crystals in 1D and in subsection \ref{prethermal} on  prethermal discrete time crystals
in arbitrary dimension and with power-law interactions.

\subsection{Many-body localized discrete time crystal}
\label{MBLDTC}

Our focus in this subsection will be on highlighting the key features of time crystalline order in Floquet MBL systems, by describing recent observations from three distinct experimental platforms: trapped atomic ions \cite{yao2017discrete,zhang2017observation},  spins in solid state materials \cite{abobeih2019atomic,randall2021observation}, and  superconducting qubits \cite{arute2019quantum,mi2021observation,frey2022realization}.

Before describing the experiments, we begin by introducing a standard model for a period-doubled discrete time crystal in a spin-1/2 chain, and discuss several limits of this model. 
While the experiments will not implement this specific Floquet unitary, the spirit of their Floquet dynamics will be captured by this example. 
In particular, consider Floquet time evolution (with a period, $T_0=t_1+t_2$) governed by alternating between two  time-independent Hamiltonians:
\begin{equation}
H(t) = 
\begin{cases}
{H_1}\, , &\text{for time } t_1\\
{H_2}\, , &\text{for time }  t_2.
\end{cases}
\label{MBLham1}
\end{equation}
Let us take $H_1$, $H_2$ as given by:
\begin{align}
H_1 &= -\sum_{\langle i,j\rangle} J_{ij} \sigma^z_i \sigma^z_{j} -
\sum_i \left( h^z_i \sigma^z_i + h^y_i \sigma^y_i + h^x_i \sigma^x_i \right)\cr
H_2 &= g \sum_i \sigma^x_i.
\label{MBLham2}
\end{align}
with $\vec{\sigma}$ being Pauli spin operators and  $J_{ij}, h^x_i, h^y_i, h^z_i$  sufficiently disordered to ensure that the spin chain is many-body localized. 
The simplest way to see the (trivial) emergence of a period-doubled (i.e.~$\nu = 1/2$ subharmonic) response is to consider the decoupled limit ($J_{ij}=0$) with only a longitudinal field, $h_z$, along $\hat{z}$. For any individual spin initially along the $\hat{z}$-axis, the spin will Larmor precess around the $\hat{x}$-field during the second portion, $H_2$, of the Floquet evolution. When timed appropriately, $t_2 = \pi/g$, this evolution implements a so-called $\pi$-pulse, which flips the spin, and causes a period-doubled response. 
However, much like our discussion of the map $\Phi(x) =  -x$ in Section~\ref{sec:what_defines}, this period doubling is not rigid to perturbations and not indicative of a many-body phase. 
In particular, any imperfections in the timing of the $\pi$-pulse will immediately lead to the breakdown of the period-doubled response~\cite{yao2017discrete}. 
However, by turning on the Ising interactions between the spins, in conjunction with the disorder which leads to MBL,  the system becomes robust to small imperfections and the subharmonic response is rigid to arbitrary, weak perturbations of both the initial state and the Hamiltonian (so long as these perturbations respect the period of the Floquet evolution). 
We note that this last parenthetical distinguishes time crystalline order in closed systems from time crystalline order in open systems, where perturbations that explicitly break the discrete time translation symmetry of the dynamics can still be allowed (see Section~\ref{sec:toom}).

\subsection{Experimental signatures of disordered time crystals}
\label{expt_MBLDTC}

\subsubsection{Trapped ion spin chains} 
\label{MBL_DTC_ions}

\onlinecite{zhang2017observation} observed and characterized  time crystalline behavior  in a spin chain (Fig.~\ref{fig:original_expts}a) composed of trapped atomic ions. 
In the experiment,  each ion acts as an effective spin, and interactions are controlled through external, optical spin-dependent forces \cite{Monroe:2021}. 
The system is periodically driven by successively applying three time-independent Hamiltonians, corresponding to a  global drive, interactions,  and disorder~\cite{smith2016many}. 
The disorder is programmed by individually addressing each spin with a tunable laser beam. 
The ion chain was composed of a relatively small number of effective spins (between $L=10-14$) and the time crystalline response exhibited some sensitivity to the initial conditions \cite{zhang2017observation, khemani2019brief}, suggesting that long-range interactions could be leading to a time crystalline response stabilized by prethermalization (Sec.~\ref{prethermal}) rather than localization. 
The experiment was able to observe a cross-over between the DTC regime and the symmetry unbroken phase (Fig.~\ref{fig:original_expts}b).
To do so, \onlinecite{zhang2017observation} measured the Fourier spectrum of each individual spin and studied the variance of the amplitude of the $\nu = 1/2$ subharmonic response as a function of the $\pi$-pulse imperfection, $\varepsilon$. 
By increasing the strength of the interactions between the ion spins, they demonstrated that the location of the variance peak shifted toward larger values of $\varepsilon$, consistent with the expectation that many-body interactions are essential for stabilizing time crystalline order.

\subsubsection{Spins in condensed matter} 
\label{MBL_DTC_diamond}

Signatures of time crystalline order in solid-state systems have been observed in both disordered Nitrogen-Vacancy (NV) ensembles~\cite{choi2017observation,ho2017critical} and NMR systems ~\cite{rovny2018observation,rovny2018p,pal2018temporal,luitz2020prethermalization}. In the context of NV centers (Fig.~\ref{fig:original_expts}c,d), despite the presence of strong disorder, it is known that dipolar interactions in three-dimensions cannot lead to localization; however, the resulting time-crystalline order can  exhibit an anomalously long lifetime owing to critically slow thermalization~\cite{ho2017critical}. 
One of the puzzles arising from NMR experiments on phosphorus nuclear spins in ammonium dihydrogen phosphate was the observation of period doubling despite the lack of disorder and a high temperature initial state; these observations were ultimately understood as consequences of an approximate long-lived $U(1)$ conservation law~\cite{luitz2020prethermalization}, which leads to effective prethermal time crystalline order (see Sec.~\ref{prethermal}). 

\begin{figure*}
\centering
\includegraphics[width=1.0\linewidth]{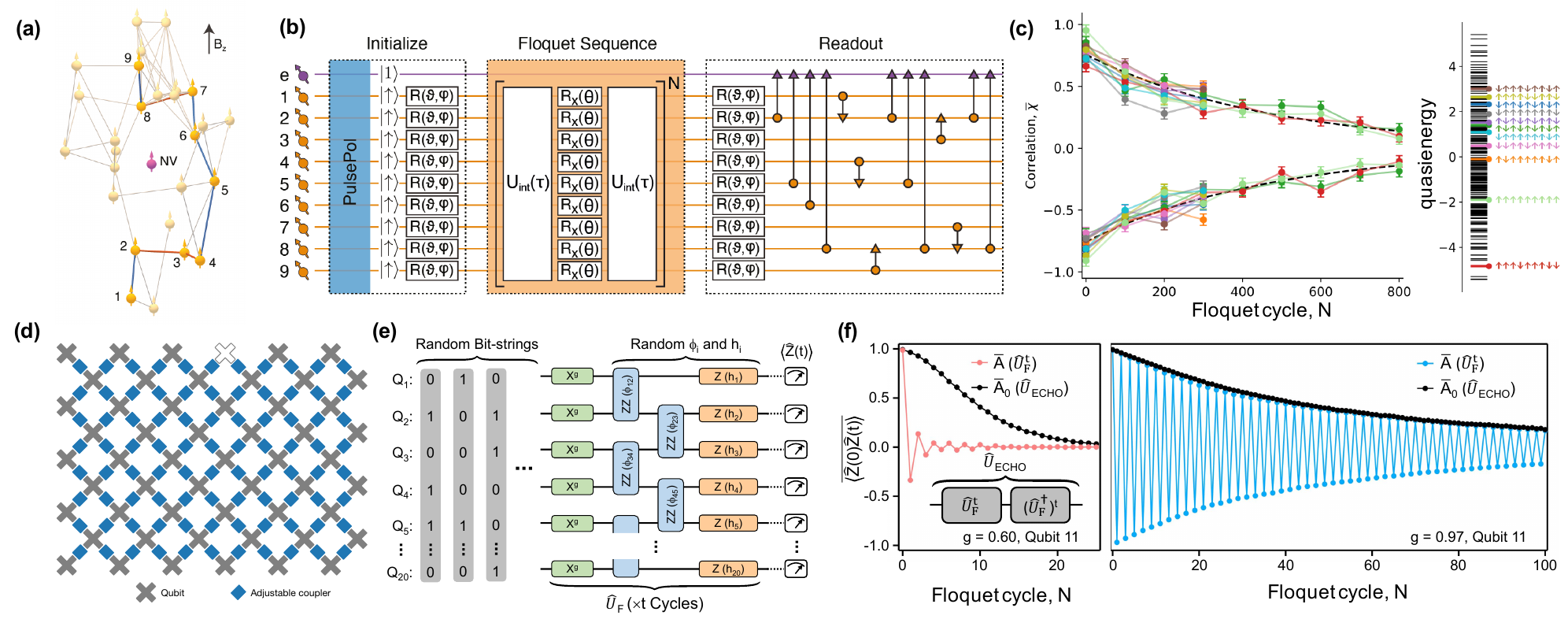}
\caption{(a) Schematic image depicting the 9 spin subset of the 27 pre-characterized nuclear spins surrounding the single NV center. (b) Each nuclear spin can be individually initialized and read-out via coupling to the NV center. The Floquet sequence consists of $U_\textrm{int}$, which includes a disordered long-ranged Ising interaction and a disordered longitudinal field followed by an approximate $\pi$-pulse, $R_x(\theta)$. The disorder is naturally inherited from the random positioning of the nuclear spins within the diamond lattice. 
(c) In the MBL discrete time crystal phase, the system exhibits robust subharmonic oscillations lasting $\sim 10^3$ Floquet cycles. The different colors are associated with different initial states, which are characterized by their quasienergy in the right panel.
Crucially, the time crystalline behavior is independent of the initial state, as expected for an MBL time crystal. 
(d) Schematic depicting the 53 transmon qubit Sycamore chip. \cite{mi2021observation} isolate one dimensional chains of $L = 8,12,16,20$ qubits from this two-dimensional grid. (e) A digital Floquet sequence consisting of random longitudinal fields, disordered nearest-neighbor Ising interactions, and approximate $\pi$-pulses are applied to random ``bit-string'' initial states. (f) In the thermal phase (left panel), the disorder averaged autocorrelation function quickly decays to zero. In the MBL time crystal phase, the same correlation function exhibits subharmonic oscillations lasting $\sim 10^2$ Floquet cycles. \cite{mi2021observation} also implement a benchmarking ``echo'' sequence, $U_\textrm{echo}$, which demonstrates that the time crystal's lifetime is consistent with being limited by experimental imperfections and decoherence. 
Figures a-c are adapted from \cite{randall2021observation}, while figures d-f are adapted from \cite{arute2019quantum} and \cite{mi2021observation}.
} 
\label{fig:C13MBLDTC}
\end{figure*}

Recent work in the solid-state has focused on pushing toward a realization of time crystalline order in a regime compatible with many-body localization~\cite{randall2021observation}. 
As depicted in Fig.~\ref{fig:C13MBLDTC}(a), \onlinecite{randall2021observation} utilize a platform consisting of a precisely characterized array of 27 nuclear spins surrounding a single NV center in diamond~\cite{abobeih2019atomic}. Owing to differences in the hyperfine interaction strengths, each nuclear spin is individually addressable (for both initialization and read out [Fig.~\ref{fig:C13MBLDTC}(b)]). 
In order to work with an effective one-dimensional geometry (thus avoiding issues regarding the stability of MBL in $d>1$), \cite{randall2021observation} select a specific 9-spin subset of the 27 nuclear spins.
By preparing a variety of product initial states,  \cite{randall2021observation} demonstrates that dipolar interactions between the nuclear spins lead to robust period doubling for nearly $\sim10^3$ Floquet cycles (note that each Floquet cycle lasts $10$~ms), independent of the initial state [Fig.~\ref{fig:C13MBLDTC}(c)]. 
Moreover, they show that local thermalization occurs within $\sim 10$ Floquet cycles, demonstrating that the observed time crystalline order is not a result of slow thermalization.

\subsubsection{Superconducting transmon qubits} 
\label{MBL_DTC_SC}

Signatures of time crystalline order have also recently been observed in three  experiments utilizing superconducting transmon qubits~\cite{mi2021observation,xu2021realizing,frey2022realization}. 
Working with Google's `Sycamore' processor, \cite{mi2021observation} employ a quantum-circuit-based approach  to demonstrating MBL time crystalline order~\cite{ippoliti2020many}. 
In particular, in order to work with a one dimensional system,  \cite{mi2021observation} isolate a nearest-neighbor coupled chain of $L = 20$ qubits from the two-dimensional array (Fig.~\ref{fig:C13MBLDTC}d); an analogous strategy is taken by \cite{frey2022realization}, who isolate an $L=57$ qubit chain on  IBM's quantum processors: \emph{ibmq\_manhattan} and \emph{ibmq\_brooklyn}.
Both \cite{mi2021observation} and \cite{frey2022realization} utilize one and two-qubit gates in order to implement a digital Floquet sequence.
For the experiment performed on the Sycamore processor, the specific Floquet sequence was given by
(Fig.~\ref{fig:C13MBLDTC}e),
\begin{equation}
U_F = e^{-\frac{i}{2}\sum_i h_i \sigma^z_i} e^{-\frac{i}{4}\sum_i J_i \sigma^z_i \sigma^z_{i+1}} e^{-\frac{i}{2} \pi g \sum_i \sigma^x_i},
\label{eq:sycamore}
\end{equation}
 consisting of  random longitudinal fields, disordered nearest-neighbor Ising interactions, and approximate $\pi$-pulses.
 Deviations from the ideal $\pi$-pulse limit (i.e.~$g=1$ in Eq.~\ref{eq:sycamore}) are used to control the transition between the discrete time crystal phase (i.e.~$g=0.97$ in Fig.~\ref{fig:C13MBLDTC}f) and the trivial thermal phase (i.e.~$g=0.6$ in Fig.~\ref{fig:C13MBLDTC}f). 
 To demonstrate that the observed subharmonic response is not affected 
  by the choice of initial states, both \cite{mi2021observation} and \cite{frey2022realization} probe the Floquet dynamics starting from  random initial bit-strings.
  In addition, \cite{mi2021observation} also experimentally implement a finite-size-scaling  analysis by varying the length of their one dimensional chain between
contiguous subsets of 8, 12, and 16 transmon qubits; this reveals a transition between the MBL discrete time crystal and the thermal phase at a critical value of the $\pi$-pulse imperfections, $0.83 \gtrsim g_c \lesssim 0.88$.
Finally, in order to characterize the intrinsic gate errors and decoherence of their system (with $T_1, T_2 \sim 100 \mu$s \cite{arute2019quantum,kjaergaard2020superconducting}), \cite{mi2021observation} implement a benchmarking ``echo'' Floquet sequence, which reverses the digital time evolution after a set number of Floquet cycles. 
Dividing by this benchmarking sequence leads to signatures of time crystalline order that exhibit minimal effective decay.

\subsection{Prethermal discrete time crystal}
\label{prethermal}

\subsubsection{Floquet prethermalization}
While many-body localization provides a quantum approach to realizing a time crystal with an infinite lifetime in the thermodynamic limit, $\tau \sim e^L \rightarrow \infty$, in this section, we focus on an alternate, disorder-free strategy, dubbed Floquet prethermalization (Fig.~\ref{fig:closed_system}).
Here, the lifetime of the time crystalline order does not scale with the system size, but can be exponentially long in a particular control parameter, namely, the ratio of the driving frequency to local energy scales, $J$, within the system: $\tau \sim e^{\omega_D / J}$.

\begin{figure}
\centering
\includegraphics[width=1.05\columnwidth]{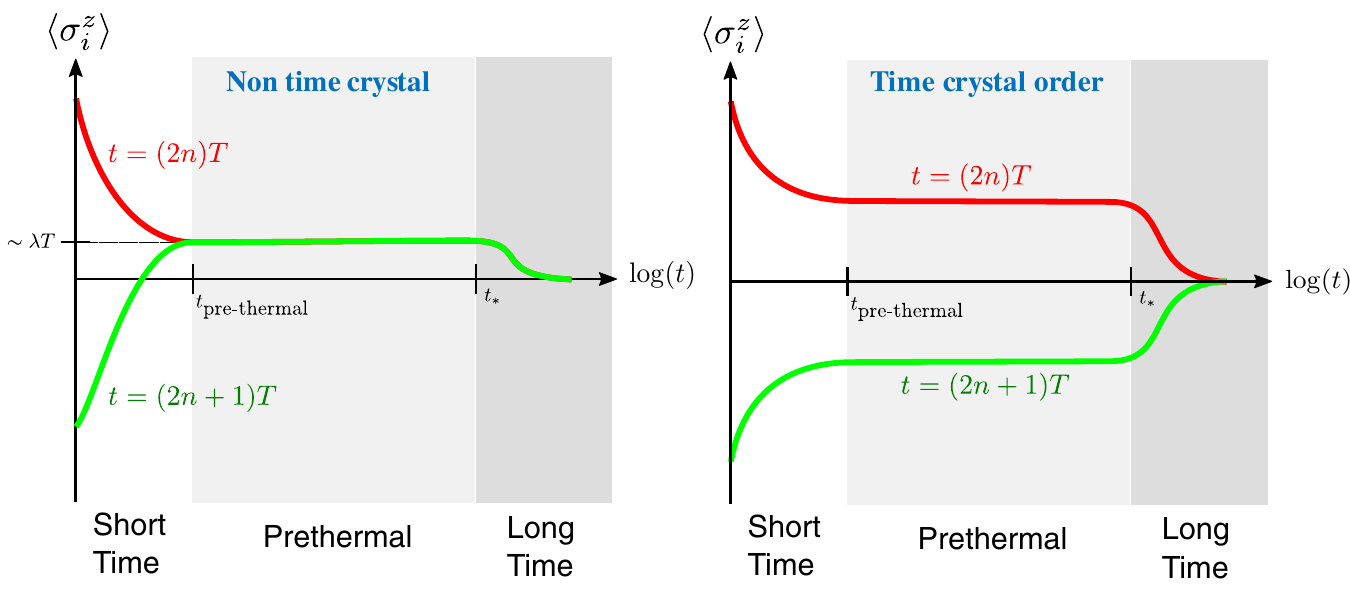}
\caption{Schematic depicting prethermal time crystalline order identified via the stroboscopic behavior of the magnetization as a function of time. (a) In the non-time-crystal prethermal phase, the magnetization during even and odd Floquet cycles becomes identical after a short time (corresponding to local thermalization). During the intermediate prethermal regime, this magnetization may remain small, but non-zero, until it decays to zero at the Floquet heating time-scale (long time). (b) In the prethermal discrete time crystal phase, the magnetization oscillates throughout the prethermal regime, leading to stroboscopic curves that exhibit plateau behavior. The prethermal time crystal ultimately melts at exponentially late times, controlled by the frequency of the driving field. 
Figure adapted from \cite{else2017prethermal}.
} 
\label{fig:prethermal_schematic}
\end{figure}

In static systems, the concept of prethermalization is a powerful framework for understanding the thermalization of systems with disparate energy scales---the dynamics (and thus thermalization) of  ``fast moving'' degrees of freedom are governed by an effective Hamiltonian where the slow degrees of freedom  remain basically frozen~\cite{kagan1974theory,berges2004prethermalization,gring2012relaxation}.
In driven Floquet systems, the presence of two distinct energy scales is particularly natural, since the frequency of the drive and the interaction energy-scales within the Hamiltonian are independent. 
Indeed, it has recently been established that Floquet prethermalization is a generic feature of driven systems in the high-frequency regime~\cite{abanin2015exponentially,mori2016rigorous,kuwahara2016floquet,abanin2017rigorous,abanin2017effective,else2017prethermal,weidinger2017floquet}.
From the perspective of  heating, as discussed in Sec.~\ref{Floquet}, one of the central consequences of Floquet prethermalization is that the frequency of the drive exponentially controls the heating time scale, $t_*$ \cite{abanin2015exponentially,mori2016rigorous,machado2019exponentially}.
The physical intuition for this exponential scaling is as follow: at large frequencies, $\omega_D \gg J$ (where $J$ is the local energy scale of the many-body system), the system must undergo $\sim \omega_D / J$ local rearrangements in order to absorb a single unit of energy from the drive. 
This intuition holds for both quantum and classical systems, although in the remainder of this subsection, we will focus on the quantum setting; we will return to the classical case in the outlook (Sec.~\ref{CPDTC}).

Crucially, \cite{kuwahara2016floquet} and \cite{abanin2017rigorous}  proved that up until the timescale, $t_* \sim e^{\omega_D/J}$, the system does not absorb energy from the drive and the stroboscopic  Floquet dynamics are (up to exponentially small corrections) captured by an effective static Hamiltonian,  $H_{\mathrm{eff}}$.
Unlike MBL, these features of Floquet prethermalization
are largely independent of disorder, dimension and the range of interactions~\cite{machado2019exponentially,machado2020long,rubio2020floquet, peng2021floquet,pizzi2021higher,fan2020discrete}.

While the ability to exponentially delay the onset of Floquet heating is crucial for realizing and stabilizing a discrete time crystal,
one additional key insight is still needed, namely, the observation that $H_{\mathrm{eff}}$ can exhibit an emergent symmetry (which need not be present in the original Floquet evolution) protected by the discrete time-translation symmetry of the drive~\cite{else2017prethermal,machado2020long}.
The presence of such an emergent symmetry in $H_{\mathrm{eff}}$ allows one to sharply define phases of matter (i.e.~via symmetry-breaking) in the prethermal regime (Fig.~\ref{fig:prethermal_schematic}).
A mathematical description of prethermal discrete time crystals in the
quantum setting closely follows the discussion of sections \ref{sec:rotating}
and \ref{sec:internal}.

This lays the foundation for the connection to time crystalline order---if a many-body system prethermalizes to a state that spontaneously breaks the emergent symmetry of $H_{\mathrm{eff}}$, it will also exhibit time-crystalline order, corresponding to a subharmonic oscillation between the different symmetry sectors~\cite{else2017prethermal,machado2020long}.

To recap, within the framework of Floquet prethermalization, realizing a time crystal requires  a few ingredients:
first, the Floquet drive must induce an emergent symmetry in the prethermal effective Hamiltonian $H_{\mathrm{eff}}$. Second, 
$H_{\mathrm{eff}}$ must be able to host a symmetry-broken phase with respect to the emergent symmetry. Note that owing to Landau-Peierls-type arguments, this naturally places constraints on the interaction range and dimensionality for realizing a prethermal discrete time crystal (PDTC)~\cite{machado2020long,kyprianidis2021observation,pizzi2021higher}. %
Finally, the initial state of the many-body system must have a sufficiently low energy density (measured with respect to $H_{\mathrm{eff}}$), such that it equilibrates to the spontaneously symmetry broken phase  during the prethermal regime (Fig.~\ref{fig:prethermal_schematic}). 

\subsubsection{
Prethermal discrete time crystal in a 1D trapped ion chain}
\label{longrange}

To highlight the dynamical signatures of a prethermal discrete time crystal, as well as the distinctions from an MBL time crystal (Sec.~\ref{MBL_DTC_diamond} and Sec.~\ref{MBL_DTC_SC}), we turn to a recent experiment performed on a trapped ion quantum simulator~\cite{kyprianidis2021observation}.
As discussed in Sec.~\ref{MBL_DTC_ions}, some of the first experimental observations of time crystalline behavior (Fig.~\ref{fig:original_expts}a,b) were originally observed in small-scale trapped ion experiments~\cite{zhang2017observation}, and one of the central advances in recent work is the ability to experimentally distinguish between local thermalization (e.g.~``short time'' in Fig.~\ref{fig:prethermal_schematic}) and late-time dynamics in the prethermal regime. 

\begin{figure}
\centering
\includegraphics[width=0.7\columnwidth]{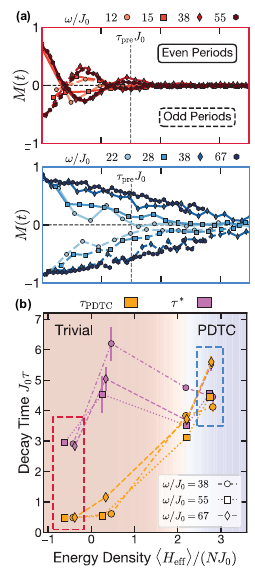}
\caption{Prethermal discrete time crystal in a disorder-free chain of $L=25$ $^{171}$Yb$^+$ ions. (a) For initial states whose energy density (measured with respect to $H_\textrm{eff}$) lies in the trivial phase of the effective Hamiltonian  (top panel), the dynamics of the magnetization is largely independent of the drive frequency and quickly equilibrates to zero. Despite the rapid decay of the magnetization, the energy density exhibits slow Floquet heating. For initial states whose energy density lies in the symmetry-broken phase of $H_\textrm{eff}$ (bottom panel), the stroboscopic dynamics of the magnetization exhibit subharmonic oscillations (solid curves correspond to even periods while dashed curves correspond to odd periods), whose lifetime extends with increasing frequency. The lifetime of this prethermal time crystalline order is consistent with being cutoff by the slow Floquet heating time-scale. 
(b) Experimentally measured phase diagram of a prethermal discrete time crystal as a function of the energy density of the initial state.  States near the edge of the spectrum (i.e.~which order with respect to $H_\textrm{eff}$) exhibit PDTC order while states in the trivial phase of $H_\textrm{eff}$ do not exhibit prethermal time crystalline order. Figures adapted from \cite{kyprianidis2021observation}.} 
\label{fig:PDTC_ions}
\end{figure}

The experiment consists of a one dimensional chain of $N=25$ Ytterbium ions and the Floquet evolution alternates between two types of  dynamics.
First, a global $\pi$-pulse is applied and then second, the system evolves under a disorder-less, long-range, mixed-field Ising model.
At leading order in the Floquet-Magnus expansion, the stroboscopic dynamics of the ions are captured by the effective Hamiltonian,
\begin{align}
H_\textrm{eff} = \sum_{i,j} J_{ij} \sigma^x_i \sigma^x_{j} + B_y
\sum_i   \sigma^y_i,
\label{Heff_ions}
\end{align}
which exhibits an emergent Ising symmetry. 
To begin, \cite{kyprianidis2021observation} demonstrate that independent of the initial state of the ion chain, the system exhibits slow, frequency-dependent heating to infinite temperature. 
As aforementioned, in order for the system to exhibit time crystalline order, the initial state of the ion chain must reach a pseudo-equilibrium state (within the prethermal regime) in which it breaks the emergent Ising symmetry (Fig.~\ref{fig:PDTC_ions}).
For a one dimensional system at finite temperature, this is only possible for sufficiently long-ranged interactions~\cite{thouless1969long,fisher1972critical,kosterlitz1976phase}.
The trapped ion system generates such long-range interactions by using a pair of Raman laser beams to couple the internal spin states to motional modes of the ion chain~\cite{sorensen1999quantum}.

Initializing the system with a  N\'{e}el state (top, Fig.~\ref{fig:PDTC_ions}a), \cite{kyprianidis2021observation} observe that the magnetization, $M(t) = 1/N \sum_{i=1}^N \langle \sigma^x_i(t) \rangle \langle \sigma^x_i(0) \rangle$ quickly decays to zero, in agreement with the expectation that the system equilibrates to a symmetry- unbroken paramagnetic state.
On the other hand, starting from a  polarized initial state, $M(t)$ exhibits period doubling (bottom, Fig.~\ref{fig:PDTC_ions}a), whose lifetime  is directly controlled by the frequency of the drive. 
Moreover, in this latter case,  the lifetime of the time-crystalline order matches with $t_*$, consistent with the intuition that Floquet heating ultimately melts the PDTC at late times.

An intriguing open question is whether this melting can be delayed or fully arrested by coupling the system to a
cold bath~\cite{else2017prethermal}.
More broadly, the stability and rigidity of many-body time crystalline order in dissipative, open quantum systems [Fig.~\ref{fig:bec_time_crystal}(c)] remains an active area of  exploration~\cite{iemini2018boundary,zhu2019dicke,lazarides2020time,booker2020non,buvca2019non,kessler2020observation,kessler2020continuous,lledo2019driven,gambetta2019discrete,seibold2020dissipative,chinzei2020time,lledo2020dissipative,gong2018discrete,dogra2019dissipation,buvca2019dissipation,tucker2018shattered,kessler2019emergent,kongkhambut2022observation}. We return to it in Section \ref{sec:open}.

\subsection{Periodically driven  Bose-Einstein Condensates \label{sec:driven_BEC}}
\begin{figure}
\centering
\includegraphics[width=0.9\columnwidth]{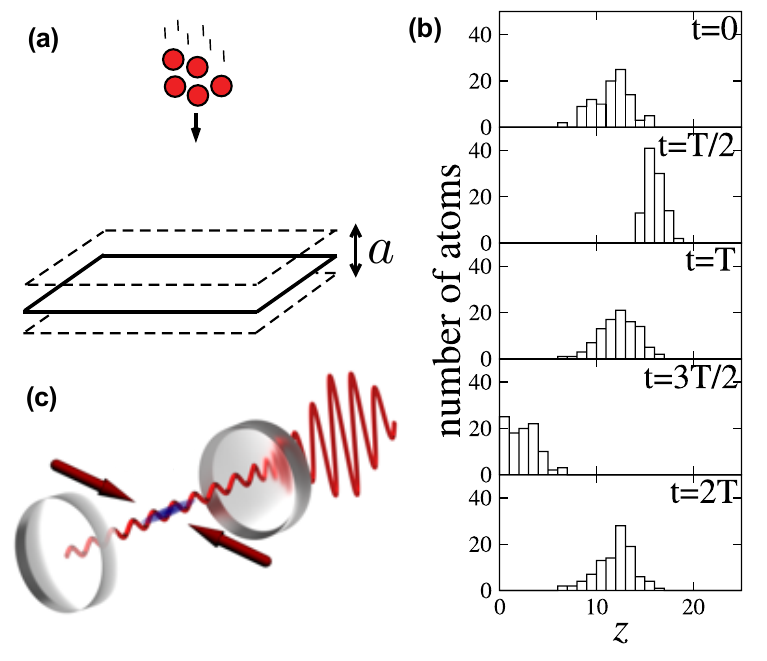}
\caption{(a) Schematic depicting $N$ atoms bouncing off an oscillating mirror~\cite{sacha2015modeling}. (b) Numerical simulations of the scenario in (a) shows that the time evolution of the atomic density  exhibits \TTSB{}. At each time shown, the positions of $10^2$ atoms (out of $10^4$) are measured and a histogram of those positions is depicted. (c) A Bose-Einstein condensate in a transversely pumped high-finesse optical cavity can realize signatures of \TTSB{} in an open system.
In particular, the atom-cavity system can exhibit a period-doubled oscillation between
distinct density wave patterns~\cite{kessler2020observation}.
Figures adapted from \cite{sacha2015modeling,sacha2017time,kessler2020observation}.} 
\label{fig:bec_time_crystal}
\end{figure}

Long-lived subharmonic responses have  been explored extensively in  driven Bose Einstein condensates (BEC)  without disorder~\cite{sacha2015modeling,smits2018observation,autti2018observation,giergiel2020creating, liao2019dynamics,smits2020long, smits2021spontaneous, autti2021ac,wang2021discrete,sacha2020time,hannaford2022condensed,kongkhambut2022observation}.
A paradigmatic example is provided by a BEC bouncing off an oscillating mirror \cite{flatte1996classical, sacha2015modeling}.
Let us try to understand this scenario by considering a particle under the influence of gravity which bounces off an oscillating mirror [Fig.~\ref{fig:bec_time_crystal}(a)]. 
In the reference frame of the mirror (in which the mirror is held constant at $z=0$), the particle's gravitational potential energy is given by  $U(z) = m (g  + a \cos(\omega_D t)) z$, with the position of the particle at $z > 0$.
Classically, a single particle in this potential exhibits a 2:1 subharmonic response which is conceptually equivalent to that of the driven non-linear pendulum discussed in Sec.~\ref{sec:floq_H_emergent} \cite{flatte1996classical}.
In the quantum case, this leads to two ``non-spreading wavepackets'' localized to the double-minima of the effective Floquet Hamiltonian, corresponding to the two period-doubled orbits   ~\cite{bialynicki1994lagrange,buchleitner1995nondispersive,flatte1996classical, buchleitner2002non}.
However, in contrast to the classical case, quantum tunneling between the two minima results in a finite lifetime, $\tau$, for the subharmonic oscillations.
From the spectral perspective [Sec.~\ref{sec:spectral_perspective}], this manifests as two Floquet eigenstates, $|\psi_{1} \rangle$ and $|\psi_{2} \rangle$,   whose relative Floquet eigenvalues are perturbed away from $\pi \rightarrow \pi + 1/\tau$.
Intuitively, these two Floquet eigenstates correspond to the 
 ``bonding'' and ``anti-bonding'' configurations of the effective double-well potential.
When $\tau \to \infty$, the configurations $|\psi_{\pm} \rangle= \frac{1}{\sqrt{2}} (|\psi_1\rangle \pm |\psi_2\rangle)$ are exchanged under each period, and  a system prepared with a majority amplitude in one or the other will thus exhibit \TTSB{}. 
But with tunneling, these configurations slowly precess into each other over the tunneling time $\tau$, restoring the symmetry~\cite{holthaus1994subharmonic, holthaus1995classical, flatte1996classical, buchleitner2002non, sacha2015modeling}.

However, $\tau$ can be radically extended by replacing a single particle with an $N$-particle  Bose-Einstein condensate with an attractive contact interaction $H_{\textrm{int}} = -\frac{g_0}{4} |\Psi(z)|^4$~\cite{sacha2015modeling}.
In appropriate units, the Hamiltonian for the field operator $\hat{\Psi}$ is given by
\begin{align}
\hat{H} &= \int_0^\infty dz \left[ \hat{\Psi}^\dagger(z) \left(-\frac{1}{2}\partial_z^2 + (1 + \frac{a}{g} \cos(\omega_D t))|z| \right)\hat{\Psi}( z)  \right. \nonumber \\ 
&\quad \left. -\frac{g_0}{4} |\hat{\Psi}(z)|^4 \right ].
\end{align}
Two approximations can be used to analyze the resulting dynamics~\cite{sacha2015modeling}: (i) taking $N \to \infty$ while holding $g_0 N$ fixed, the  dynamics reduce to a periodically driven Gross-Pitaevskii equation (GPE).  The GPE is a classical Hamiltonian field theory over a complex field $\psi(t, z)$, and can therefore be efficiently simulated; (ii) one can project the interacting dynamics onto the two nearly-degenerate ground states (i.e.~the aforementioned double-minima) of the single particle Floquet Hamiltonian, effectively reducing the problem to two bosonic modes.

These approximations can be combined into three cases.
When both approximations are made (i.e.~taking  $N \to \infty$ and also projecting to the two-mode picture), the dynamics exhibit a bifurcation transition that produces period-doubling.\cite{sacha2015modeling}
Intuitively, due to the attractive interaction, the Floquet energy is minimized when all particles inhabit the same well; tunneling between the wells  is thereby suppressed because all $N \to \infty$ bosons must tunnel sequentially. 

If one only considers the second approximation (i.e.~finite $N$ within the two mode picture), quantum fluctuations in the mode occupation remain, and it was shown that the tunneling time increases exponentially with $N$, $\tau \propto e^{\alpha N}$~\cite{sacha2015modeling, wang2021discrete}.
Thus,  TTSB is achieved not in the limit of large system size, but rather in the limit of large particle number.
However, the two-mode approximation rules out  the Floquet heating that is generally presumed to occur in a driven many-body system ~\cite{dalessio2014longtime,khemani2019brief,else2020discrete}.

Finally, if one only considers the first approximation (i.e.~$N \to \infty$ but without taking the two-mode limit), \TTSB{} effectively requires that the periodically driven Gross-Pitaevskii equation is non-ergodic. 
This is in contrast to the  expectation that  classical non-linear field theories are generically ergodic (the integrability of the isotropic, undriven GPE being a fine-tuned exception). 
However,  this expectation may not apply to the present situation. 
Due to the gravitational confining potential, the dynamics are effectively confined to a zero-dimensional region near the minima of the double-well.
This is in contrast to the Gross-Pitaevskii equation in an isotropic potential or an array of coupled pendula, which each have an \emph{extensive} density of states. 

So, while the number of degrees of freedom is formally infinite, it may be that the KAM theorem  still applies because there are effectively only a few degrees of freedom that are  resonant with the periodic drive.
Thus, KAM might
protect \TTSB{} in this scenario as it does for a single particle in a double-well Floquet potential (Sec.~\ref{sec:shaken_pendulum}). %
Interestingly, in the absence of periodic driving, this is indeed the case; specifically, it has been proven that the  undriven GPE in a double-well potential breaks ergodicity. Below a critical energy, a condensate becomes  ``self-trapped'' in one of the minima~\cite{jackson2004geometric, albiez2005direct}.
It is perhaps then plausible that the driven GPE can exhibit the same phenomena, but in the rotating frame of the 2:1 resonance [Fig.~\ref{fig:bec_time_crystal}(b)].
Numerical simulations of the driven GPE support this interpretation~\cite{sacha2015modeling, kuros2020phase}.

Finally, there is of course the question of what occurs when one makes neither of the two aforementioned approximations. 
 Exact answers are difficult to obtain in this regime, as it is a full quantum many-body problem, but numerical results have been obtained within  time-dependent Bogoliubov theory
 ~\cite{kuros2020phase} and the truncated Wigner approximation~\cite{wang2021many}.
Both conclude that out to thousands of driving periods, the boson population remains almost entirely within the two modes, with no evidence for heating. 

These findings support the conclusion of the two-mode approximation (with finite $N$), where $\tau \propto e^{\alpha N}$.
Of course such numerical simulations cannot definitively rule out exponentially slow depopulation and heating (Sec.~\ref{prethermal}), as might occur in a prethermal time-crystal~\cite{else2017prethermal,sacha2020time,machado2019exponentially}. 
However, if one accepts the conclusion that as $N \to \infty$,  the GPE can generically break ergodicity in this setting (as it does in a \emph{static} double well potential), the possibility that Floquet heating is absent out to infinite times even for the driven GPE, seems like a reasonable (albeit, still surprising) possibility.

Finally we comment on the contention that time-crystals in driven BECs are ``effectively few-body''~\cite{khemani2019brief}.  This is partly a matter of nomenclature. On the one hand, these systems are clearly not few-body as the interaction, $g_0$, and the limit $N \to \infty$ are necessary for  \TTSB{}. Furthermore, while not rigorously proven (much like  Floquet MBL is also not rigorously proven), the authors of this Colloquium find it plausible that \TTSB{} could be stable up to times, $\tau \propto e^{\alpha N}$, even when the physics is treated as a full many-body problem.
Again, much like the situation with MBL, this ``plausibility'' is bolstered by the fact that the undriven GPE in a double-well potential does, in fact, break ergodicity.

On the other hand, it seems that  \TTSB{} is only stable because of the reduced dimensionality induced by the gravitational confinement, which effectively ensures the accuracy of the two-mode approximation;  no such picture would be obtainable for the MBL setting (Sec.~\ref{MBLDTC}). 
We venture that a more useful distinction is the nature of the thermodynamic limit being taken. 
In the case of MBL time-crystals (or the driven, open time-crystals of Sec.~\ref{sec:open}), true \TTSB{} is achieved in the thermodynamic limit $L \to \infty$, with $\tau \sim e^{L/\xi}$, keeping intensive quantities fixed. 
In the case of the driven BEC, due to the confining gravitational potential, the system size $L$ is not relevant (in this sense the problem is zero dimensional) and the problem reduces to being effectively few-mode, even though the physics is realized in a full, many-body system. 
In this setting, \TTSB{} is recovered in the limit $N \to \infty$, keeping $g_0 N$ fixed, with $\tau \sim e^{\alpha N}$.

An intriguing probe of this distinction would be to consider a 2D generalization of the driven BEC:  bosons
     are gravitationally confined in the $z$ direction, but  propagate freely along $x$.
     In this case, the density of states is extensive in $x$, and domain walls between the two period-doubled orbits can form. 
     It is natural to suppose that  \TTSB{} is prethermal in this case, as could be evaluated in the $N \to \infty$ limit using the 2D GPE. 

\section{Open, periodically-driven systems and stochastic dynamics}
\label{sec:open}

Our discussion thus far has focused on closed systems in which the dynamics are deterministic, $x \to \Phi(x)$.
However, when a system is coupled to an environment, it is often fruitful to model the effect of the environment's chaotic motion  as noise (most-conveniently taken to be Markovian), so that the dynamics become effectively stochastic. Rather than focusing on a particular microstate, one instead considers a probability distribution over microstates $\rho(x, t)$  which evolves under a  ``master equation,'' for example the Fokker-Planck equation.
Integrating the master equation over one Floquet cycle then produces a discrete update of the distribution, $\rho(x, t + T) = \Phi[\rho(x, t)]$.
In  classical mechanics this results in a probability distribution over canonical coordinates $\rho(x = \{p, q\}, t)$ which evolves under a Markov process, $\rho(x', t + T)  = \int \Phi(x' | x) \rho(x, t) dx$, where $\Phi(x' | x)$ are the Markov transition probabilities.
In the quantum case, we have  a density matrix $\hat{\rho}(t)$ which evolves under a ``quantum channel'' $\hat{\rho} \to \Phi[\hat{\rho}]$.
One can also in principle consider non-Markovian baths, a point to which we will return. 

    The definition of \TTSB{} given in Eq.~\eqref{eq:def_TTSB} generalizes to the open case by measuring the local observable, $O$, in expectation, and stability can be defined by requiring \TTSB{} be robust to  perturbations of the stochastic dynamics subject to locality and any other dynamical constraints one is interested in [Fig.~\ref{fig:bec_time_crystal}(c)]. 
    The environment is both good and bad for \TTSB{}.
    On the one hand, coupling to an environment introduces friction: the energy and entropy produced by the periodic drive can now be absorbed by the bath, which can prevent the long-time Floquet heating which would otherwise destroy \TTSB{} in the absence of MBL. This tends to help stabilize   spontaneous time translation symmetry breaking.
    On the other hand, if the environment is at finite temperature,  the inevitable noise which results may occasionally conspire to cause phase slips in the period-doubled motion. 
    Roughly speaking, if noise nucleates phase slips at rate $1/\tau$, the \TTSB{} has a finite auto-correlation time $\tau$ and there is no true long-range order. 
    The interplay of a periodic drive, interactions, dissipation, and noise results in a truly non-equilibrium situation which is exceptionally rich -- just like the world around us~\cite{iemini2018boundary,zhu2019dicke,lazarides2020time,booker2020non,buvca2019non,kessler2020observation,kessler2020continuous,lledo2019driven,gambetta2019discrete,seibold2020dissipative,chinzei2020time,lledo2020dissipative,gong2018discrete,dogra2019dissipation,buvca2019dissipation,tucker2018shattered,kessler2019emergent,kongkhambut2022observation}.

    \subsection{``Activated'' time-crystals}
    \label{sec:activated_tc}
A classical  realization of this interplay is given by  \emph{Langevin dynamics}.
Due to the coupling with the environment, each degree of freedom $(q, p)$ experiences an additional Langevin force  $F_L(t) = - \eta \dot{q} + f(t)$. Here $\eta$ is the friction coefficient and $f(t)$ is a white-noise stochastic force with  auto-correlation function  $\langle f(t) f(t') \rangle = 2 \eta T \delta(t - t')$, where $T$ is the temperature of the bath. The total force is obtained by adding $F_L$ to the time-dependent Hamilton's equations, $\dot{p} = -\partial_q H(t) + F_L(t)$. 
This leads to a master equation -- the Fokker-Planck equation -- of the general form $\partial_t \rho(\{p, q\}, t) = \mathcal{L}[\rho]$, where $\mathcal{L}$ is the Fokker-Planck operator.
Integrating the Fokker-Planck equation over one Floquet cycle of the the drive then gives the discrete-time update $\rho(t + T) = \Phi[\rho(t)]$.

Note that we are assuming  the environment remains in equilibrium at temperature $T$, and so instantaneously satisfies the fluctuation-dissipation theorem which relates the magnitude of the friction $\eta$ and the noise $2 \eta T$. If $H(t)$ were time-independent, this would ensure (via detailed-balance) that the system  relaxes to the canonical ensemble at long-times where \TTSB{} is forbidden~\cite{bruno2013impossibility, nozieres2013time,watanabe2015absence}.
But when $H(t)$ is periodically driven, this  need not be the case: \emph{a non-equilibrium steady state can develop in which energy and entropy flows from  the drive to the bath via the system}.

\begin{figure}
\centering
\includegraphics[width=0.75\columnwidth]{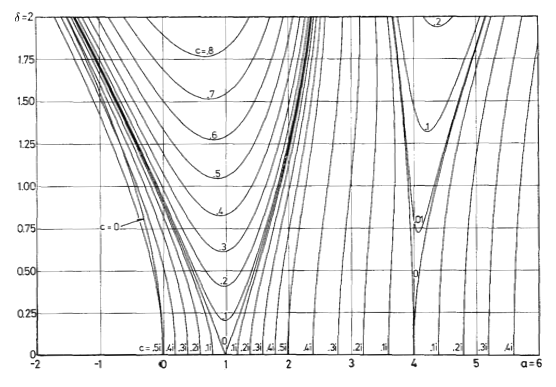}
\caption{Arnold  tongues of the damped Matthieu equation. The $x$-axis is the drive frequency expressed via the dimensionless ratio $a = (2 \omega_0)^2 /  \omega^2_D$, while the $y$-axis is the drive amplitude $\delta$. Contours indicate the critical drive amplitude required for resonance when the coefficient of friction takes a given value $c$.
When $c > 0$, the period-double solution ($a = 1$) onsets only at finite $\delta$.
Figure adapted from \onlinecite{pedersen1980stability}.
\label{fig:pederson_stability}}
\end{figure}

We can now return to the question of  period-doubling in open classical many-body systems such as the  Faraday-wave instability and coupled pendula (see e.g.~earlier discussions surrounding Eq.~\eqref{eq:H_FK} and Sec.~\ref{sec:faraday_waves}).
For small oscillations and weak damping, the friction results in a \emph{damped} Mathieu equation of the general form $\ddot{q}_k = - \left[\omega_k^2  + \delta_k \cos(\omega_D t) \right] q_k - \gamma_k \dot{q}_k$. 
The sub-harmonic  responses of this model have been studied extensively \cite{pedersen1935subharmonics,hayashi1953subharmonic,taylor1969stability,pedersen1980stability}.
As shown in Fig.~\ref{fig:pederson_stability}, for $\gamma_k > 0$ the subharmonic response  acquires a finite threshold value  for the drive amplitude $\delta_k$, and modes outside the resonance tongues are damped. 
Absent noise,  \TTSB{} is then realized as a sharp many-body bifurcation transition as some combination of driving frequency, amplitude  or damping strength is tuned into the $2$:$1$ parametric resonance of a $k$-mode. 
For strongly viscous Faraday waves, the equations of motion are somewhat more complex, but the conclusion are similar~\cite{edwards1994patterns, kumar1994parametric}. 

\begin{figure}
\centering
\includegraphics[width=0.9\columnwidth]{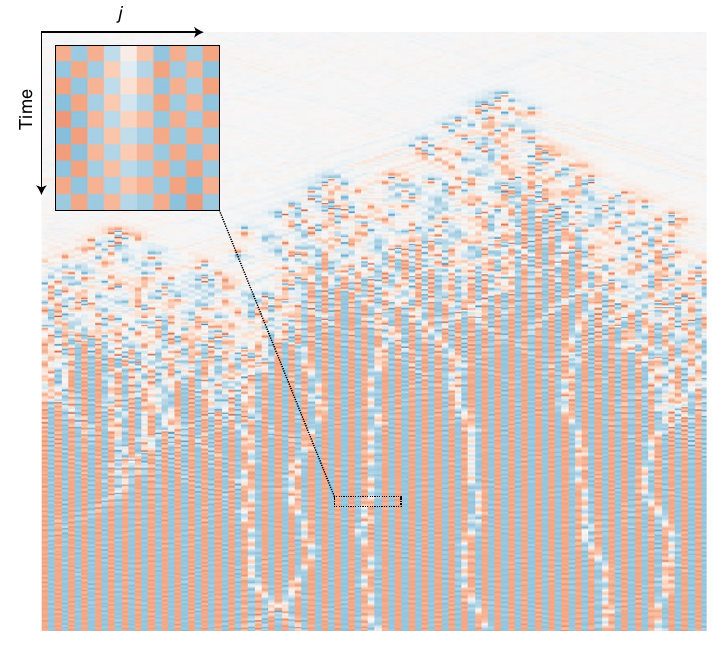}
\caption{Domain walls between different period-doubled solutions of the parametrically driven 1D Frenkel-Kontorova model (Eq.\eqref{eq:H_FK})  in contact with a finite-temperature Langevin bath.
Colors indicate the amplitude of the $j$-th oscillator observed at even stroboscopic times $q_j(t = 2 n T)$, with time running vertically. Inset, the oscillations at stroboscopic times $q_j(n T)$, which reveal the period doubling. 
Figure adapted from \cite{yao2020classical}.
} 
\label{fig:domain_wall_CDTC}
\end{figure}

However,  noise (i.e.~finite temperatures, $T>0$)  induces fluctuations between the different cycles of the \TTSB{}, leading to a finite auto-correlation time.
Recall from our discussion of prethermal systems (Sec.~\ref{sec:rotating} and Sec.~\ref{prethermal}) that we may transform to a rotating frame $\mathcal{K}(t)$:~$H(q, p, t) \to H(P, Q, t) = H_{\textrm{eff}}(Q, P) + V(t)$ in which the driving $V(t)$ is weak.
For a single pendulum (Eq.~\ref{eq:H_pendulum}), $H_{\textrm{eff}}$ takes the form of a double-well potential. 
The noise will then lead to activated hopping between the minima at rate $1/\tau = e^{-\Delta / k_B T}$, where $\Delta$ is the quasi-energy barrier between the two minima.
Thus, at finite temperature, a single pendulum is no longer a true time-crystal, but rather an ``activated'' time-crystal in which the autocorrelation time of the \TTSB{} diverges exponentially with temperature.

The problem is richer when coupling the pendula into an array, where $H_{\textrm{eff}}$ approximately takes the form of an  Ising model (recall Sec.~\ref{sec:shaken_pendulum} for e.g.~the definition of $P_*$),
\begin{align}
H_{\textrm{eff}} \approx \sum_i \left[  \frac{(P_i - P_\ast)^2}{2} +  a  \cos(2 Q_i) \right] +  \frac{\tilde{g}}{2} \sum_{\langle i, j \rangle}  (Q_i - Q_j)^2,  
\label{eq:coupled_pendula_array}
\end{align}
where the parameters $a$ depends on $\{\delta, \epsilon, \omega_D/\omega_0\}$ (from Eq.~\ref{eq:H_FK}) and $\tilde{g}$ is the transformed coupling strength.

The low quasi-energy configurations are then localized \emph{domain walls} between the different period-$m$ orbits.
These domain walls are easily observed when numerically simulating a 1D coupled array of pendula [Eq.~\eqref{eq:coupled_pendula_array}] in the presence of the Langevin noise, as shown in Fig.~\ref{fig:domain_wall_CDTC}. 
Indeed, one observes a gas of domain walls  undergoing Brownian motion due to the noise, occasionally nucleating or annihilating in pairs.
\cite{yao2020classical} studied the motion of these domain walls in detail and found that it leads to a decay in the \TTSB{} autocorrelation function, $\langle q_i(t = n T) q_i(t=0) \rangle \sim (-1)^n e^{-t/ \tau}$.
The decay rate $\tau$ is proportional to the density of domain walls, which is in turn related to the temperature of the bath through an Arrhenius law, $\tau \sim e^{\Delta / k_B T}$, where $\Delta$ is now the quasi-energy activation barrier required to nucleate a domain wall. 

 \cite{yao2020classical} found numerically that in 1D, collective effects can cause the barrier $\Delta$ to drop discontinuously as the parameters (e.g.~$\delta$ or $\omega_D$) are varied, indicating a non-equilibrium phase transition at which \TTSB{} is completely destroyed. 
Without careful examination of the temperature dependence at low $T$, this transition might easily be mistaken as evidence for the existence of a ``true'' \TTSB{} phase ensconced within a non-equilibrium phase transition, while in reality the low temperature behavior is always activated.

Naively, the existence of a finite critical temperature, $T_c$, for equilibrium Ising symmetry breaking in dimensions two and greater suggests that the \TTSB{} domain walls cannot proliferate for $T < T_c$, which would thereby stabilize time crystalline order to infinite times.
However, in the damped system this is \emph{not} correct. 
To see why, \cite{yao2020classical} evaluated the effect of the Langevin force in the rotating frame  and found two contributions.
First, the $\{Q, P\}$ experience a  Langevin force at the same temperature $T$, which locally equilibriates $H_{\textrm{eff}}$, explaining the activated processes discussed above. 
Second, the effective Hamiltonian acquires a new contribution from the friction $\eta$, of the form $\sum_i \eta P_\ast Q_i$.
Its physical origin is that in the rotating frame, $P$ encodes the amplitude of the oscillation (cf.~Eq.~\eqref{eq:K_approx}), and because the friction prefers to damp the amplitude, this equates to a net force on $Q$.
The net potential, $V \approx a \cos(2 Q) + \eta P_\ast Q$, then biases $Q$ to ``roll downhill'', which following Eq.~\eqref{eq:K_approx}, would unlock the response from $\omega_D/2$.
The bias leads to a net force on the domain walls which causes  nucleated islands to grow~\cite{yao2020classical}, even though both sides of the domain wall are related by the $\mathbb{Z}_2$ time-translation symmetry.
This net force destroys the possibility of phase coexistence. 
Thus, even in 2D and greater, the time-crystalline response in this and similar models still exhibits  an activated auto-correlation time.
Intriguingly, a closely related effect has been studied in the context of probabilistic cellular automata,~\cite{Bennett1990Stability} which are introduced in Sec.~\ref{sec:PCA}.

As $\eta \to 0$, the magnitude of the aforementioned force goes to zero and the density of domain walls can then undergo a transition analogous to the equilibrium Ising  transition. 
However, once $\eta \to 0$, one needs to again worry about heating from the residual time-dependent part of the drive (this is precisely related to the prethermal situation discussed in Sec.~\ref{prethermal}).

In summary, the existence of $H_{\textrm{eff}}$ provides a unifying framework for three regimes of time-crystalline behavior: in the closed cased ($\eta = T = 0$), one has the prethermal scenario where heating destroys \TTSB{} after an exponentially long time $\tau \sim e^{\omega_D / J}$; in the open case ($\eta > 0, T>0$), the environment nucleates domain walls which destroy the \TTSB{} on an exponentially activated time-scale $\tau \sim e^{\Delta / k_B T}$, where $\Delta$ is a quasi-energy barrier; and finally for the purely dissipative case  $\eta > 0, T = 0$, there is true \TTSB{} conceptually analogous to that of the coupled  map lattice (Sec.~\ref{sec:illustrative}).

\subsection{Experimental realizations of activated time-crystals: pendula,  AC-driven charge density waves, and fractional Shapiro steps}
\label{sec:ac_CDW}

    Several experiments on open many-body systems have observed subharmonic responses which may be interpreted within the  framework of activated \TTSB{}.
    One obvious example is a  shaken pendulum treated as a  macroscopic object of $\sim 10^{23}$ particles.
    The coupled oscillator array of Eq.~\eqref{eq:H_FK}, for example, can be understood as a macroscopic pendulum composed of atoms, $q_i$, when accounting  only for its 1D width. 
    The effect of noise on such systems is of practical interest; for example  parametric resonance of a mesoscopic mechanical oscillator such as an AFM tip \cite{rugar1991mechanical} is a standard tool for mass and force sensing. 
    In perfect isolation,  driving will generate stresses on the oscillator, causing the phonon modes ($q_{k\neq 0}$) to absorb a portion of the driving power, heating up the system and (at long times) melting the oscillator. 
    This is the prethermal scenario, though some work remains to rigorously prove there is a regime in which the time-scale increases exponentially~\cite{mori2018floquet,ye2021classical}.
   Contact with a bath (e.g., air at $T = 293$K) prevents the oscillator from melting, and it is helpful to estimate the resulting activation time.
    In particular, the nucleation of a ``domain wall'' corresponds to ripping the oscillator in half, an energy scale so many orders of magnitude above $T = 293$K, that the time-scale is inconceivable.
    However, collisions with an unlikely conspiracy of air molecules can also cause the entire oscillator to collectively hop between the two quasi-energy minima; using Eq.~\eqref{eq:H_FK}, one can estimate  the quasi-energy barrier for such a fluctuation as  $\Delta \sim 2 \delta(2 - \omega_D / \omega_0)$.

        At scales of kilograms, meters, and room-temperature, $\Delta / k_{\textrm{B}}  T$ again results in  an inconceivably long time-scale. However, for  mesoscopic resonators it can be highly relevant.
        An application of this effect is found in nanoscale mass sensing based on parametric oscillation of an AFM tip.
        In order to weigh an object attached to the tip, one detects the sharp jump in the amplitude of oscillations  as the detuning $2 - \omega_D / \omega_0$ is driven near parametric resonance; the precise location of the jump is highly sensitive to the total mass.
        However, finite-$T$ noise rounds out the transition into a non-linear crossover, an effect which is experimentally observable and limits the practical resolution of parametric-resonance-based  sensing~\cite{zhang2004noise, prakash2012parametric}.

    Two more intrinsically mesoscopic examples, which can be recast in the language of activated time-crystalline order,  are  AC-driven charge density waves (CDW) and fractional Shapiro steps in Josephson junction arrays. 
    Both have received tremendous experimental attention; since the two systems are conceptually equivalent \cite{macdonald1983study, bohr1984transition},  we will largely focus on the former.
    In a charge density wave material such as NbSe$_3$ \cite{gruner1988dynamics}, the electron density spontaneously develops a charge density modulation at twice the Fermi wavevector, $n(x) \sim n_0 \cos(2 k_F x + \theta(x))$, where $\theta$ is the slowly varying phase of the CDW.
     Treating $\theta$ as a single macroscopic degree of freedom, one obtains a phenomenological equation of motion   given by \cite{gruner1981nonlinear}: 
\begin{align}
\ddot{\theta} + (\omega_0 \tau)^{-1} \dot{\theta} + \sin \theta = E(t) / E_T
\label{eq:cdw_eom}
\end{align}
where $\sin \theta$ accounts for the  potential that pins the CDW, $\omega_0 \tau$ is a relaxation time (note that time is rescaled by the natural oscillation frequency $\omega_0$) due to dissipation from the material, $E(t)$ is the applied bias, and $E_T$ is the threshold bias for transport. 
This is the equation of motion for a damped and driven pendulum.
When $E(t)$ is larger than the threshold $E_T$, the system enters into a sliding state with $\langle \partial_t \theta \rangle \neq 0$, generating a finite current.

\begin{figure}
\centering
\includegraphics[width=1.0\columnwidth]{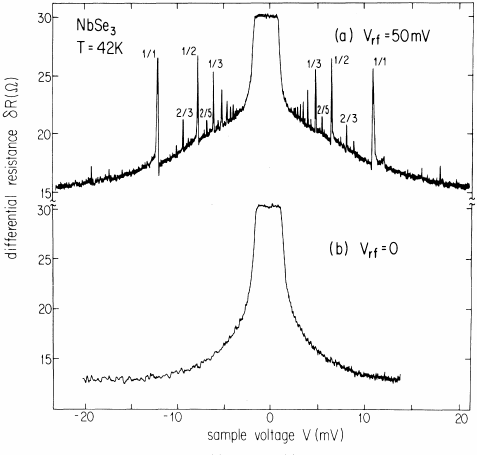}
\caption{Differential resistance measurements, $dV/dI$, in the charge density wave system NbSe$_3$. A bias voltage $V(t) = V_{\textrm{dc}} + V_{\textrm{rf}} \cos(\omega_D t)$ is applied, and the DC component of the resistance $dV_{\textrm{dc}}/dI_{\textrm{dc}}$ is measured, holding $V_{\textrm{rf}}$ and  
$\omega_D$ fixed. Sharp peaks in the resistance correspond to plateaus in the $I$-$V$ curve. 
The location of these peaks can be attributed to a motion of the CDW in which it shifts by $p/q$ wavevectors per driving period, resulting in a subharmonic response between the AC components of the voltage and current. 
In the absence of the AC drive (panel b), the subharmonic response is absent. Figure adapted from \cite{brown1984subharmonic}.
} 
\label{fig:ac_driven_cdw}
\end{figure}

In the experiments of interest, a bias is applied with both DC and AC components, $E(t) = E_{\textrm{dc}} + E_{\textrm{ac}} \cos(\omega_D t)$.
In the sliding state, the motion of the CDW over the periodic pinning potential defines a frequency scale
$\omega_{\textrm{dc}} \equiv  \langle\dot{\theta} \rangle \approx \omega_0 \tau \frac{E_{\textrm{dc}}}{E_T} $. 
When $\omega_{\textrm{dc}} = \frac{p}{q} \omega_D$,  commensuration effects allow the AC-oscillations to assist the motion of $\theta$ over the potential.
Since the current $I_{\textrm{dc}}$ is proportional to  $\omega_{\textrm{dc}}$, commensuration  can lead to a plateau in  $I_{\textrm{dc}}$ whenever $\omega_{\textrm{dc}} \approx \frac{p}{q} \omega_D$.
This leads to the emergence of a  ``devil's staircase''  of plateaus~\cite{zettl1983observation, brown1984subharmonic, thorne1987charge} in the $I$-$V$ curve as  $E_{\textrm{dc}}$ is swept at fixed $E_{\textrm{ac}}$ (see Fig.~\ref{fig:ac_driven_cdw} which depicts $dV/dI$ versus the voltage $V$).

Most notably, for $q \neq 1$ (see for example the prominent peak in Fig.~\ref{fig:ac_driven_cdw} for $p/q = 1/2$), the response is subharmonic: the CDW shifts by $1/q$ of a wavelength with each period of the AC-drive.

While Eq.~\eqref{eq:cdw_eom} treats $\theta$ as single macroscopic variable, in the experiment there are spatial fluctuations which can be accounted for by considering a 2D or 3D array of $\theta(\mathbf{r})$ coupled through an elastic stiffness.
\cite{middleton1992complete} showed theoretically that the subharmonic response remains robust in this many-body setting.
In fact, in the overdamped regime with a purely sinusoidal pinning potential, subharmonic responses are \emph{only} found when treating the problem as many-body one.

The phenomenological equation of motion [Eq.~\eqref{eq:cdw_eom}], which predicts perfect \TTSB{}, also neglects the effect of  thermal fluctuations that are of course present in  experiments such as Fig.~\eqref{fig:ac_driven_cdw}.
As discussed in \cite{thorne1987chargeI, thorne1987charge}, a careful examination of the experimental $I$-$V$ curves reveals that some (but not all) of the subharmonic plateaus are  rounded-out in a sample dependent fashion.
Complementary to this, the spectral distribution  $I(\omega)$ (or $V(\omega)$ in current biased mode) shows peaks at $\omega = \frac{p}{q} \omega_D$ with finite width, rather than perfect Bragg peaks. 
This broadening is attributed to a distribution of velocities $\partial_t \theta$ throughout the sample, which, in the language of time-crystalline order, implies the absence of perfect \TTSB{}.
Indeed, numerical simulations of driven 1D CDWs confirm that finite temperature broadens the plateaus~\cite{mali2012effects}. 
Unfortunately, owing to thermal gradients generated due to Ohmic heating, AC-driven CDWs  do not appear to be particularly well suited for quantitatively investigating the temperature dependence of the broadening at low $T$ (see e.g.~discussions in Sec.~\ref{sec:activated_tc}).

Mathematically equivalent physics is found in the AC Josephson effect (recall related discussions in Sec.~\ref{intro}). In the  McCumber model \cite{mccumber1968effect} for a resistively and capacitively shunted Josephson junction (RCSJ), the superconducting phase difference $\phi$ across a Josephson junction obeys the equation of motion
\begin{align}
{\omega_p^{-2}}\ddot{\phi} + {\omega_c^{-1}} \dot{\phi} + \sin \phi = I(t) / I_c
\label{eq:JJ_eom}
\end{align}
where ${\omega_p}\equiv \sqrt{2 I_c/C}$,
${\omega_c}\equiv 2 I_c R_N$, and $V/{R_N}=\dot{\phi}/2{R_N}$ is the normal
current across the junction; here, $C$ is the capacitance, $I$ is the current across the Josephson junction, $I_c$ is the current, and $R_N$ is the normal resistance. The situation considered in Sec.~\ref{intro}
occurs in experiments in which
the junction is under-damped.
The DC current is carried entirely
by normal quasiparticles, $I_{DC}={I_n}$, satisfying $I_{DC}/{I_c} = {\omega_c^{-1}}\dot{\phi}$ or, simply,
$I_{DC}=V/R_N$.
An AC supercurrent flows in response
to the voltage drop across the junction, $I_{AC}(t)/{I_c}
= \sin(2I_{DC}{R_N}t)$. However,
in this under-damped regime, the
$I-V$ curve is hysteretic, and
there is a branch in which
the voltage drop across the junction is zero so long as
the DC current is less than the critical current: $I_{DC}={I_c}\sin\phi$. This vertical step in the $I_V$ curve
is followed by a plateau at $I=I_c$ from $V=0$ to $V={I_c}{R_N}$ where
it rejoins the other branch.

Returning to the general case of
Eq. (\ref{eq:JJ_eom}), we note that
this equation of motion is
formally equivalent to the CDW [Eq.~\eqref{eq:cdw_eom}], and by analogy one also expects steps in the $I$-$V$ curve when the AC component of the current, $I(t) = I_{\textrm{dc}} + I_{\textrm{ac}} \cos(\omega_D t)$, is commensurate with rational harmonics of the DC  voltage, $\omega_J = \langle \dot{\phi} \rangle = 2 e V_{\textrm{dc}}$.
(The vertical step in the superconducting
branch of the hysteretic $I-V$
curve in the underdamped case is a trivial
version of such a step.)
These ``Shapiro steps'' were first observed at integer harmonics in single Josephson junctions~\cite{shapiro1963josephson}. Later experiments~\cite{benz1990fractional, lee1991subharmonic} on arrays of Josephson junctions revealed subharmonic Shapiro steps.

\subsection{Ergodicity in open systems}
\label{sec:open_ergodicity}
     At time scales $t > e^{\Delta / k_B T}$, an activated time crystal loses memory of the initial condition which distinguishes between the $m$-cycles of the \TTSB{} oscillation: its dynamics are ultimately ergodic. 
      Earlier in this Colloquium, ergodicity was introduced as a property of a measure-preserving deterministic system.
The precise definition works differently in the stochastic case, but intuitively, still captures whether a system inevitably ``forgets'' its initial condition.
We direct the interested reader to \cite{gielis2000coupled, gray2001reader} for a discussion of  the technical aspects required to make the definitions precise in the thermodynamic limit.

In the stochastic setting, a probability distribution $\rho_s$ is an  ``invariant measure'' of the dynamics if it is a fixed point of the stochastic update,  $ \Phi[\rho_s] = \rho_s$, i.e.~it is a steady-state distribution. 
A stochastic system is said to be ergodic if two properties hold: (1) the dynamics have a \emph{unique} steady state $\rho_{s}$ and (2) the  long-time behavior of any initial state relaxes to this steady state, $\lim_{t \to \infty} \rho(t) = \rho_s$.
As in the closed, deterministic case, \TTSB{} in the sense of Eq.~\eqref{eq:def_TTSB} requires that $\Phi$ is non-ergodic.
Otherwise, at long times, the $m$-possible orbits  become indistinguishable and any oscillations will decay.

Instead, a \TTSB{}  phase  will exhibit so-called ``asymptotic periodicity''\cite{lasota1984asymptotic, losson1996thermodynamic}.
At long times the distribution relaxes to a convex combination of $m$ locally-distinguishable distributions $\rho_p$,  $\rho(t) \to \sum_{p=1}^m \alpha_p \, \rho_{p+t}$  which are cyclically permuted under the evolution $\Phi[\rho_p] = \rho_{p+1}$ (with $\rho_{p+m} = \rho_p$).
While $\rho_s = \frac{1}{m} \sum_p \rho_p$ is a unique steady state, in the generic case where the $\alpha_p$ are unequal $\rho(t)$ will  continue to oscillate  so that the limit $\lim_{t \to \infty} \rho(t) = \rho_s$ fails to exist~\footnote{In a deterministic system, ``ergodicity'' governs the invariance of Ces\'aro sums,  $\lim_{\tau \to \infty} \frac{1}{\tau} \sum^\tau_{t = 1} O(\Phi^{(t)}(x)) $.  In this case,  \TTSB{} implies that the $m$-fold iterated map $\Phi^{(m)}$ is not ergodic, but the map  $\Phi$ itself generally \emph{is} ergodic (though it is not mixing). In the stochastic case, however, ergodicity is usually defined to imply the stronger form of convergence $\lim_{\tau \to \infty} = \rho_s$, rather than $\lim_{\tau \to \infty} \frac{1}{\tau} \sum^\tau_{t = 1}  \Phi^{(t)}[\rho_0]  = \rho_s$. By this definition, $\Phi$ \emph{itself} is non-ergodic. 
While the steady state $\rho_s = \frac{1}{m} \sum^m_{p=1} \rho_p$ is unique, and one has  $\lim_{\tau \to \infty} \frac{1}{\tau} \sum^\tau_{t = 1}  \Phi^{(t)}[\rho_0]  = \rho_s$,  the limit $\lim_{\tau \to \infty} \rho(t)$ does not exist.
}.

The existence of  true \TTSB{} in open dynamical systems thus hinges on a far more fundamental question: can a locally interacting stochastic system \emph{generically} break ergodicity? This line of questioning has a deep history in the fields of non-linear dynamics, mathematical physics, and computer science. 
The answer is intimately related to the stability of \emph{phase coexistence} and \emph{phase transitions} in stochastic systems.

We can illustrate the ideas at play by  stripping away the complexity of Hamiltonian dynamics and instead considering a classical spin system. 
For specificity we start in equilibrium with a classical Ising model $H_{\textrm{I}}[\{\sigma\}]$.
While $H_{\textrm{I}}$ itself does not define any dynamics, we may define a ``kinetic Ising model,'' such as Glauber or Hasting-Metropolis rules, which update the spin configuration $ \sigma  \to \sigma'$  according to a local conditional probability $\Phi( \sigma' | \sigma)$ which obeys detailed balance with respect to $H_{\textrm{I}}$. 
Detailed balance ensures that the distribution  $\rho_s = e^{-\beta H} / Z$ is a steady state.
On a \emph{finite} system at $T > 0$, standard results  guarantee that $\rho_s$ is the unique steady state with a finite relaxation time, so the process is ergodic~\cite{feller1957introduction}. 
However, the thermodynamic limit leads to richer possibilities.
When $h  = 0$ and $T$ is below the critical Ising temperature (e.g., when tuned to a first-order phase transition),  the phase coexistence of the two magnetized phases implies the existence of \emph{two} steady states, $\rho(\sigma)_+ = \rho(-\sigma)_-$, and the dynamics are not ergodic.  
However, for any finite $h$ phase coexistence is lost and the steady state is unique.
This illustrates a general principle in equilibrium: ergodicity-breaking requires tuning one or more parameters to a first-order phase transition, and in this sense is fine-tuned (or requires a symmetry beyond time-translation invariance). 

We note that one route around this  equilibrium no-go is to consider a spin Hamiltonian with power-law interaction of sufficiently slow fall-off, so that the energy penalty for domain walls can be made to grow faster than the extensive energy of the symmetry-breaking external field~ \cite{liggett2012interacting}.
Or, taking this to the extreme, by considering models with all-to-all couplings in the spirit of a mean-field Hamiltonian, so that locality is lost entirely~\cite{morita2006collective,russomanno2017floquet,pizzi2021bistability,yang2021dynamical,lyu2020eternal}. 
However, such models are not local in the usual sense:  their thermodynamic properties are not extensive in system size, among other anomalies.
So, while such models can be used as the basis for stabilizing even continuous \TTSB{}  \cite{kozin2019quantum}, the starting point is rather different than the usual notion of locality implied in the classification of phases.

\subsection{Probabilistic cellular automata}
\label{sec:PCA}
Moving beyond equilibrium, one can consider more general local Markov updates $\Phi(\sigma' | \sigma)$ of a discrete spin system without reference to any particular $H$.
Such models are a type of \emph{probabilistic cellular automata} (PCA)\cite{dobruvsin1990stochastic}; the continuous-time generalization of PCAs are dubbed  ``interacting particle systems''  \cite{liggett2012interacting}.
One particularly convenient way to obtain a PCA is to start with a deterministic cellular automata (CA) \cite{gutowitz1991cellular} defined by a local update rule $\sigma \to \Phi_{\textrm{CA}}(\sigma)$, and then  follow each CA update with random spin flips at an ``error rate'' $\varepsilon$. 
For Markovian errors we then obtain a Markov process which is a perturbation of the deterministic CA, $\Phi =  \Phi_{\varepsilon} \circ \Phi_{\textrm{CA}}$ (we note that mathematical results on PCA are even more general, allowing for non-Markovian errors).
Such models have been of interest to the theoretical computer science community because CA are Turing complete, so the study of the possible \emph{generic} behaviors of a PCA have implications for the generic behaviors obtainable by a classical computer perturbed by errors: e.g.~Can reliable systems emerge from unreliable components~\cite{von2016probabilistic}?
For a review of these connections between theoretical computer science and more familiar notions in non-equilibrium statistical physics, we refer the interested reader to \cite{gray2001reader}.

From the perspective of time crystals, the relevant question becomes the following: Do there exist CA which remain non-ergodic for \emph{generic} small perturbations $\Phi_{\varepsilon}$? And if so, might such a CA be the basis for spontaneous time translation symmetry breaking?
These questions were answered in the affirmative in \cite{toom1980} and \cite{Bennett1990Stability} 
Before diving into the first question, we can sketch Bennett's answer to the second. 
Suppose there does exist a local, non-ergodic CA  $\Phi_{\textrm{CA}}$ with $m$-distinct steady states. For the simplicity of discussion, we further assume that these states are exchanged by a $\mathbb{Z}_m$ symmetry, but (unlike a kinetic Ising or Potts  model) we assume the non-ergodicity is stable to \emph{generic} (possibly $\mathbb{Z}_m$ breaking!) errors. %
We can then define  a rotating version of the CA  by composing  $\Phi_{\textrm{CA}}$  with the action of $\mathbb{Z}_m$,  $\Phi_{\frac{2 \pi}{m}\textrm{CA}} \equiv G_{2 \pi/m} \Phi_{\textrm{CA}}$.
By construction, $\Phi_{\textrm{CA}}^m = \Phi_{\frac{2 \pi}{m}\textrm{CA}}^m$; thus, if $\Phi_{\textrm{CA}}$  is proven to be stably non-ergodic,  $\Phi_{\frac{2 \pi}{m}\textrm{CA}}$  inherits this property and by construction exhibits the ``asymptotic periodicity'' of a time-crystal, cycling through the $m$-fixed steady states of $\Phi_{\textrm{CA}}$.

A particularly important point to emphasize is that (while convenient) it is \emph{not} necessary to assume that $\Phi_{\textrm{CA}}$ exhibits a $\mathbb{Z}_m$ symmetry for stability. Indeed,  if the non-ergodicity of $\Phi_{\textrm{CA}}$ is stable to  $\mathbb{Z}_m$-breaking perturbations, one could  just as well have  started out with a perturbed version of $\Phi_{\textrm{CA}}$ which breaks the $\mathbb{Z}_m$ symmetry. 
The construction thus satisfies the important conceptual requirement that \TTSB does not depend on any further internal symmetries.

    Returning to the first question, the possibility of a local and generically non-ergodic PCA was a long-standing question for several decades~\cite{dobruvsin1990stochastic, lebowitz1990statistical, toom1995cellular}    
finally answered in the affirmative in the groundbreaking work of \cite{toom1980}.

\begin{figure}
\centering
\includegraphics[width=0.8\columnwidth]{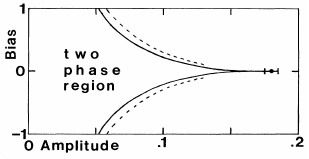}
\caption{Phase diagram of the Toom model, adapted from \cite{bennett1985role}. After each CA update, spins randomly flip up or down with rate $\varepsilon_p$, $\varepsilon_q$ respectively, with ``bias'' $(\varepsilon_p-\varepsilon_    q)/(\varepsilon_p+\varepsilon_q)$ and ``amplitude'' $\varepsilon_p+\varepsilon_q$. Within the two-phase region, there are two distinct steady-state distributions and the dynamics are not ergodic. 
Unlike the Ising model, the system can thus ``remember'' one bit of information even in the presence of \emph{biased} noise. } 
\label{fig:bennett_role}
\end{figure}

\subsection{An absolutely stable open time-crystal: the $\pi$-Toom model}
\label{sec:toom}
 The Toom CA is a 2D binary spin model with a remarkably simple ``north-east corner'' (NEC) majority rule, defined as follows~\cite{toom1974nonergodic, toom1980}.  At each step, each spin follows a majority vote amongst itself and its two north and east neighbors:  $\sigma_{x, y}(t+1) = \textrm{maj}( \sigma_{x, y}, \sigma_{x+1, y}, \sigma_{x, y+1})$.
This is a seemingly innocuous modification of the $T=0$  nearest-neighbor kinetic Ising rule, in which each spin would follow a vote amongst its four $\{$north, east, south, west$\}$ neighbors. 
Note, however, that due to the spatial asymmetry, the Toom rule \emph{cannot} be understood as minimizing the energy of any Hamiltonian, and does not obey detailed balance~\cite{grinstein1985statistical}.
This asymmetry ensures that if an island of the minority spin nucleates, the NEC-rule will cause the island to shrink linearly in time from the NE-direction inward.
This is far faster than the Ising model, where the thermodynamic force on a domain wall depends on its local curvature, and hence decays with the size of the island~\cite{bennett1985role}.

 The Toom PCA is then obtained by perturbing with \emph{biased} errors in which, for example, spins flip up  with error rate $\varepsilon_p$, and down with error rate $\varepsilon_q$. The bias $b = (\varepsilon_p - \varepsilon_q)/(\varepsilon_p+\varepsilon_q)$ breaks the Ising symmetry. 
 Remarkably, there is nevertheless a finite volume in  $\{\varepsilon_p, \varepsilon_q\}$-space in which \emph{two} magnetized steady states persist  despite the bias that prefers one over the other (Fig.~\ref{fig:bennett_role}). 
 Rigorous results prove that the ergodicity breaking is robust to  essentially \emph{any} sufficiently small perturbation --- even a spatio-temporally correlated one, or a noise distribution which is not itself time-translation invariant~\cite{toom1980, berman1988investigations, gacs2021new}.
 We note that this last point distinguishes the ``absolute stability'' of the open system $\pi$-Toom time crystal (to be defined in the next sentence)  from its closed system counterparts (i.e.~Sec.~\ref{DTCFloquet}), where noise that breaks the underlying discrete time translation symmetry would immediately destroy the time crystal.

   With Toom having done the difficult part, the $\pi$-Toom model is then defined as the \emph{anti}-majority NEC  rule $\sigma_{x, y}(t+1) = \textrm{anti-maj}( \sigma_{x, y}, \sigma_{x+1, y}, \sigma_{x, y+1})$.
   The $\pi$-Toom model is a time-crystal, which is stable to arbitrary perturbations.
  Interestingly, the \TTSB{} of the $\pi$-Toom, model was already pointed out over two decades ago in \cite{Bennett1990Stability}, and later in \onlinecite{gielis2000coupled}, where it was referred to as an example of a ``type-G phase transition.''
   
      While the original Toom model exhibits $m=2$ steady states, and hence is the basis for period $m=2$ \TTSB, it generalizes to any $m>2$  \cite{Bennett1990Stability}.
   Furthermore Toom-like models exist in all dimensions $D > 1$, simply by stacking the 2D version.
    In fact, an extension in $D = 3$  provides the basis for error-corrected classical computing~\cite{gacs1988simple}. 
   The existence of stable ergodicity breaking in 1D PCAs was only shown more recently by  G\'acs,  \onlinecite{gacs2001reliable}.
   Taken together, PCAs support absolutely stable  \TTSB{} of any period $m \geq 2$ in all dimensions $D \geq 1$.

   Is the full error-correcting capability of the Toom and G\'acs models really necessary for stable \TTSB{}? For example,  in the construction  $\Phi_{\frac{2 \pi}{m}\textrm{CA}} \equiv G_{2 \pi/m} \Phi_{\textrm{CA}}$ we may instead take $\Phi_{\textrm{CA}}$ to be the  kinetic Ising model or its generalization to $\mathbb{Z}_m$.
   The ferromagnetic interactions of the Ising model are a mild form of error correction in that they cause minority islands to shrink due surface tension.
 However, \cite{Bennett1990Stability}  argued that this equilibrium form of error correction is insufficient for stabilizing  $m > 2$ \TTSB{}.
 
   The reason is that for $m>2$, the chirality of the periodic  evolution between steady states $n \to n+1 \mod m$ implies that there is no symmetry which forbids a stochastic force \emph{per unit length} from acting  on a domain wall between  regions of $n$ and  $n+1$ - type steady states (e.g. favoring $n$ to grow at the expense of $n+1 \mod m$).
   Such a force is thus generic, and will drive minority droplets  to  nucleate and grow. 
   This force is counteracted by the surface tension of the ferromagnetism, which, however, decays with the local curvature. Thus the former force will always win out for sufficiently large droplets, causing minority droplets to proliferate and  destroy \TTSB{}  at long times.
   The Toom model escapes this reasoning because its error-correction effectively exerts a force on domain walls which is independent of their curvature, and can thus shrink minority droplets of any size.
    \cite{bennett1985role}
   
   However, the $m = 2$ case is an exception because there is no handedness to the periodic evolution. Even if there is a force which favors (say) type $n=0$ over $n=1$, because the domains effectively switch type at each step, the net force on a domain wall  averages to zero over one period, and islands can thus be shrunk via ferromagnetic surface tension~\cite{Bennett1990Stability}. 
   This implies the binary-spin $\pi$-Ising model can support \TTSB{} below the Ising critical temperature, as was recently studied in detail~\cite{gambetta2019classical}.
   
   It is interesting to contrast this discussion with the classical activated time crystal of Sec.\ref{sec:activated_tc}, which  was \emph{not} stable against the proliferation of minority islands even though $m=2$. To see why, recall that the  emergent Ising degree of freedom was effectively embedded into a larger continuous state space, e.g. $\sigma = \text{sgn}( \sin(Q))$. A domain wall between $\sigma = 1, -1$ could thus be of two types depending on whether $Q$ wound clockwise or counter-clockwise. Because this handedness is preserved under the oscillation $Q \to Q + \pi$, there is no symmetry which forbids a domain wall from experiencing a net force which does \emph{not} average to zero over $m=2$ periods, even while the identities of the domains themselves are exchanged.   
   
   Does this imply that if the binary-spin $\pi$-Ising model is embedded into a model with additional degrees of freedom, the \TTSB{} may not be generically stable? Or instead, may there be a  parameter regime of models like the Langevin-driven parametric oscillator arrays \eqref{eq:H_FK} which exhibit true, rather than activated \TTSB{}? 
   Only time will tell.

\subsection{Open Hamiltonian and quantum dynamics}
\label{sec:open_hamiltonian}
While we have discussed the existence of stable time-crystals in the stripped-down setting of a PCA (Sec.~\ref{sec:PCA} and Sec.~\ref{sec:toom}), our original microscopic starting point (Sec.~\ref{sec:open}) was the periodically driven Langevin equation, or its quantum analog, the Lindblad equation. 
An intriguing open question is the following: Do the rigorous results on \TTSB{} in probabilistic cellular automata translate to these settings?
Since PCA are  motivated as a coarse-grained description of the Hamiltonian world around us, it certainly tempting to speculate that they do.
Two recent studies have shown how the $\pi$-Toom model can be embedded into classical Langevin \cite{zhuang2021absolutely} or quantum Lindblad  \cite{mcginley2021absolutely} dynamics, and in both cases there is strong numerical evidence for true \TTSB. 
However, there is not yet a fully rigorous proof as in Toom and G\'acs work. 
While Toom-like stability for a continuous-time version of a PCA was proved in \cite{gacs2001reliable}, the state space in that model is discrete.

As we will discuss in the outlook, the answer to this question  is a special case of a deeper one:  Is error-corrected  computing \emph{physically} realizable in this universe? 

\section{Outlook and future directions}
\label{forwardT}

\subsection{New venues for time crystals}

In Sec.~\ref{ergodicity_breaking} of this Colloquium, we introduced the mantra: Where there is ergodicity breaking, there will be time-crystals. 
This naturally suggests that new  developments in the ergodicity breaking of many-body systems --- ranging from Hilbert space fragmentation~\cite{sala2020ergodicity,yang2020hilbert} and shattering~\cite{khemani2020localization} to quantum scars~\cite{turner2018weak,serbyn2021quantum} and Stark localization~\cite{schulz2019stark,vanNieuwenburg2019,doggen2021stark}  --- represents fertile ground for exploring novel formulations of time-crystalline order.

\subsubsection{Quantum many-body scars}

Interacting quantum systems can  exhibit a weak breakdown of thermalization, where certain initial conditions exhibit persistent many-body revivals in time~\cite{bernien2017probing,turner2018quantum}.
This phenomenon, dubbed quantum many-body scars, is associated with the presence of anomalous, non-thermal eigenstates; we direct the interested reader to a recent review on the topic~\cite{serbyn2021quantum}.
From the perspective of time crystalline order, the presence of scar states can cause the system to undergo periodic entanglement and disentanglement cycles following a quench~\cite{ho2019periodic,michailidis2020stabilizing,serbyn2021quantum}.
In practice, however, quantum many-body scar states  are quite fragile since they rely upon the existence of a dynamically disconnected subspace of non-thermalizing eigenstates; indeed, the presence of  generic interactions is  expected to  eventually lead to  thermalization~\cite{lin2020slow}.

The connection between quantum many-body scars and time-crystalline behavior has recently been explored in a Rydberg-based quantum simulation platform~\cite{bluvstein2021controlling,maskara2021discrete}.
In particular, the experiments observed that the coherent  revivals associated with quantum many-body scars could be  stabilized by additional periodic driving. 
This periodic driving leads to both an increase in the lifetime of the scarred oscillations as well as the emergence of a period-doubled, subharmonic response.
While reminiscent of prethermal time crystalline order (Sec.~\ref{prethermal}), we note two important differences. First, the experiments are performed in a regime where the  driving frequency is of similar order as the local energy scales of the many-body system. Second, the subharmonic response exists only for N\'eel-like initial states (associated with the quantum scars)~\cite{maskara2021discrete}.
Looking forward, a number of doors are opened by the possibility that periodic driving can  \emph{enhance} the stability of non-ergodic dynamics.
In addition to novel settings for observing time crystalline order, the ability to parametrically control the lifetime of such order could also lead to potential applications in areas such as quantum metrology and quantum information science. 

\subsubsection{Stark time crystals}

In the presence of an electric field, the wavefunction of electrons in a material are localized to a region whose size decreases as the field increases. 
In the absence of interactions, this phenomenon, which arises \emph{without} disorder, is known as Wannier-Stark localization~\cite{wannier1959elements}.
Recent theoretical and experimental investigations have explored whether such localization can persist in the presence of many-body interactions, leading to Stark MBL~\cite{schulz2019stark,guo2020stark,morong2021observation,scherg2021observing}. 

 \cite{kshetrimayum2020stark} discuss the possibility of 
 discrete time crystals protected from Floquet heating by Stark many-body localization.
They probe the existence of a Stark  time crystal using numerical simulations of a one-dimensional spin chain; focusing on two specific initial states (i.e.~the ferromagnetic and anti-ferromagnetic states), \cite{kshetrimayum2020stark} observe that for a sufficiently strong linear potential, the states exhibit time crystalline order, which seems to be robust  to $\pi$-pulse imperfections (recall discussions from Sec.~\ref{sec:MBLDTC_theory}).
 However, when differences in the linear potential  coincide with integer multiples of  the drive frequency, they observe the coherent self-destruction of  time-crystalline behavior.
An advantage of the clean, disorder-free setting, is that one can imagine the possibility of  order that spontaneously breaks both spatial and time translation symmetries. 
This  theme has been studied in a number of  contexts~\cite{trager2021real,smits2018observation,zlabys2021six} and remains an active area of exploration.

\subsection{Prethermalization beyond Floquet quantum systems}

\subsubsection{Classical prethermal discrete time crystals}
\label{CPDTC}

In Sec.~\ref{prethermal}, we focused our attention on prethermal time crystals in closed (i.e.~unitarily evolving) quantum systems. 
However, as previously discussed, one of the central features of Floquet prethermalization---i.e.~exponentially slow heating---is also expected to occur in classical many-body systems~\cite{howell2019asymptotic,mori2018floquet,hodson2020energy}. 
This immediately begs the question: Can such systems also host prethermal time crystals~\cite{howell2019asymptotic,mori2018floquet,hodson2020energy,ye2021classical,pizzi2021classical}?

Recent work by \onlinecite{ye2021classical} and \onlinecite{pizzi2021classical} answer this question in the affirmative.
A practical advantage of the generalization to classical systems, is that it immediately enables the study of PDTCs in dimensions, $d>1$.
For example, \onlinecite{ye2021classical} study a near-neighbor-interacting classical Floquet spin model on the square lattice.
Although the original Floquet evolution does not exhibit any symmetries, $H_\textrm{eff}$ exhibits an emergent $\mathbb{Z}_2$ Ising symmetry.
Unlike in the one dimensional case discussed in Sec.~\ref{longrange}, in two dimensions, this Ising symmetry can be broken at finite temperatures.

\begin{figure}[t]
\includegraphics[width=3.4in]{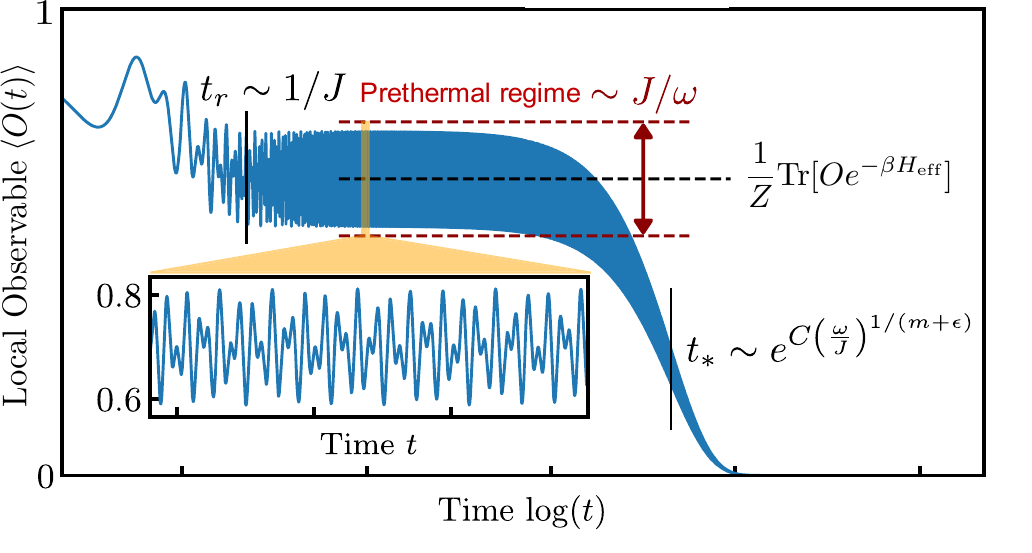}
\centering
\caption{Schematic depiction of a prethermal time quasi-crystal emerging from a quasiperiodically driven quantum system. During the prethermal regime, local observables exhibit discrete time quasicrystalline order, which ultimately melts at late times as the system Floquet heats to a featureless infinite-temperature state. Figure adapted from~\cite{else2020long}.
}
\label{fig:quasiperiodic}
\end{figure}

\subsubsection{Higher-$M$ discrete time crystals}
\label{higherM}

In both the quantum and classical settings, the emergent symmetry in $H_\textrm{eff}$ is not restricted to just a $\mathbb{Z}_2$ symmetry and different subharmonic PDTCs (i.e.~beyond  period doubling)  can be realized~\cite{giergiel2018time,surace2019floquet,pizzi2021higher,giergiel2020creating}.  
The simplest approach to doing this is to utilize a Floquet evolution where the ``$\pi$-pulse'' (see e.g. the discussion following Eq.~\ref{MBLham2}) is adjusted to be close to  a $\frac{\pi}{M}$-pulse.
In certain scenarios, this choice naturally leads to an effective Hamiltonian, $H_\textrm{eff}$, which exhibits a $\mathbb{Z}_M$ symmetry~\cite{pizzi2021classical}.
When this $\mathbb{Z}_M$ symmetry is spontaneously broken, the system will oscillate at a subharmonic frequency $\sim \omega_D/ M$ throughout the prethermal regime.
Using this strategy, \onlinecite{pizzi2021classical} construct a ``phase diagram'' for an $\omega_D/4$ PDTC in a three dimensional classical spin system. 

When one utilizes a fractional $M$ in the strategy above, the dynamics can be even richer than the integer case. 
Indeed,~\cite{pizzi2021higher} explored this scenario in the presence of long-range interactions~\cite{russomanno2017floquet,huang2018symmetry,kozin2019quantum}, observing signatures of PDTC order at a variety of fractional frequencies.

\subsubsection{Prethermal time quasi-crystals}
\label{quasiperiodically}

Building upon generalizations to higher-$M$ PDTCs and fractional PDTCs~\cite{pizzi2021higher,pizzi2021classical,ye2021classical,matus2019fractional}, a natural question to ask is whether there can be non-trivial phases of matter in isolated many-body systems that are driven in a way that is \emph{not} periodic in time. 
The focus of this outlook subsection is on the closed system, prethermal setting (Fig.~\ref{fig:closed_system}) where a periodic drive is replaced by a \emph{quasiperiodic} drive~\cite{gommers2006quasiperiodically,crowley2019topological,ringot2000experimental}. 
Recent work has shown that such systems can host prethermal quasiperiodically-driven phases of matter, and in particular can give rise to discrete time quasi-crystals~\cite{dumitrescu2018logarithmically,else2020long}.

Before jumping in however, we note that related phenomena have been explored in non-linear dissipative dynamical systems (see e.g.~the coupled map lattice discussions in Sec.~\ref{sec:illustrative})
driven at two incommensurate frequencies~\cite{sethna1984universal,held1986quasiperiodic,romeiras1987strange,ding1989evolution,flicker2018time}. 

The spectral content of a periodic drive contains peaks only at a single drive frequency, $\omega_D$, and its harmonics $n \omega_D$, for integer $n$.
By contrast, the spectral content of a quasiperiodic drive has peaks at integer linear combinations: $n_1 \omega_1 + n_2 \omega_2 + \cdots + n_m \omega_m$, for some $m > 1$ number of incommensurate frequencies,  $\{\omega_1, \cdots , \omega_m\}$.
As in  Sec.~\ref{prethermal}, in order to discuss prethermal phases, it is necessary for the system to exhibit slow-heating (i.e.~energy absorption) from the drive. 
For quasiperiodic driving, this is strictly more challenging than the original Floquet setting, since the drive is (technically) able to supply energy in arbitrary units.
Nevertheless, an analogous slow-heating result was rigorously proven by \onlinecite{else2020long}.  
Before heating occurs (Fig.~\ref{fig:quasiperiodic}), the dynamics of the quasiperiodically driven system are well-approximated (in a rotating frame) by an effective static Hamiltonian (in direct analogy to Sec.~\ref{prethermal} and Eqn.~\ref{Heff_ions}).

In order to discuss the possibility of prethermal time quasicrystals (which are stable during the prethermal time-scale, $t<t_*$), \onlinecite{else2020long} begin by defining what an  ``order parameter'' for such a phase would look like. 
The subtlety is that a quasiperiodically driven system does not, strictly speaking, have any remaining time-translation symmetry to break. 
However, such a system can still exhibit a well-defined  notion of a fractional or subharmonic frequency response (Fig.~\ref{fig:quasiperiodic}). 
In particular, an 
 observable can respond in a quasiperiodic manner with base frequencies: $\{\widetilde{\omega}_1, \cdots, \widetilde{\omega}_n\}$.
 When the $\widetilde{\omega}_j$ are \emph{not} harmonics of the original driving frequencies, then the system is said to exhibit a  fractional frequency response and discrete time quasicrystalline order~\cite{else2020long}. 
 Connecting to our previous discussions about emergent symmetries in $H_\textrm{eff}$, one can think of the discrete time quasicrystal as emerging from the spontaneous breaking of a different finite Abelian group symmetry (i.e.~which replaces the $\mathbb{Z}_M$ symmetry discussed in Secs.~\ref{sec:internal},~\ref{CPDTC}).

Finally, let us end this outlook subsection by pointing the reader to a number of related directions that fall outside the prethermal context. Many of these connect to the idea of   quasiperiodic pattern formation in parametrically driven systems. For example, in the context of two-frequency forcing, even when the frequencies are commensurate (implying that they do not exhibit a quasiperiodic response in time), in certain scenarios, one can realize stable Faraday-wave patterns that are analogous to a two-dimensional quasicrystal~\cite{edwards1993parametrically,besson1996two,silber2000two}.
Various incarnations of a quasiperiodic response in time have also been explored in both theoretical proposals and experiments on cold atomic systems~\cite{autti2018observation,giergiel2018time,giergiel2019discrete,pizzi2019period,chinzei2020time}.
As an example, \cite{giergiel2019discrete} explore the dynamics of an  ultra-cold atomic ensemble bouncing between two orthogonal harmonically oscillating mirrors. By tuning the bare  frequencies of the unperturbed particle motion, the system can reproduce fragments of the Fibonacci sequence encoded via the bounces of the atomic ensemble off of the two mirrors. %

\subsection{Applications: from metrology to quantum information benchmarking}

The periodic control underlying the discrete time crystal is, more generally, an indispensable tool in the context of nuclear magnetic resonance  spectroscopy~\cite{bloembergen1947nuclear,bloembergen1948relaxation,waugh1968approach}, quantum information science~\cite{khodjasteh2005fault,biercuk2009optimized}, and AMO-based quantum simulation~\cite{bloch2012quantum,goldman2014light}. 
Building upon this connection, \onlinecite{choi2017quantum} explored the possibility of  engineering a Floquet system, where quasienergy gaps can protect entangled states from static perturbations, while still ensuring their sensitivity to an oscillating signal.
In some sense, this idea can be understood as a generalization of spin echo spectroscopy, which utilizes many-body states; it is also related to the conventional concept of using  phases of matter with spontaneously broken symmetries for sensing~\cite{PhysRevLett.121.020402,PhysRevApplied.13.044031}.

\onlinecite{choi2017quantum} investigated the use of a time-crystalline Floquet sequence  to stabilize Schrodinger-cat states that are typically extremely fragile against local
perturbations; they analyze a technique that allows for the enhancement of metrological bandwidth, while maintaining the sensitivity and discuss an example in the context of the precision measurement of AC magnetic fields.
More recently, these ideas have also inspired new many-body driving protocols aimed at  circumventing the so-called interaction limit for quantum sensing~\cite{zhou2020quantum}. 

Beyond metrology, the fact that signatures of time crystalline order have been observed in a diverse array of physical platforms (see e.g.~Secs.~\ref{MBL_DTC_ions},~\ref{MBL_DTC_diamond},~\ref{MBL_DTC_SC}), suggests the possibility of cross platform benchmarking for the performance of near term quantum devices~\cite{preskill2018quantum}.
This idea builds on the more general thread that exploring non-equilibrium dynamical phenomena may represent a particularly natural strategy for verifying and validating noisy intermediate-scale quantum technology.
Finally, suggestions for utilizing time crystalline order as a frequency standard or for beyond-SQL (standard quantum limit) quantum sensing have also been discussed~\cite{lyu2020eternal}, although explorations along this direction remain relatively nascent. 

\subsection{Concluding Remarks}

As we have seen in this Colloquium, recent evidence points to (at least) two venues where infinitely-long-lived time crystals can exist: (i) \TTSB{} which is stable to  time-translation invariant perturbations can occur in periodically-driven closed 1D quantum systems in the presence of strong disorder (Sec.~\ref{MBLDTC}) and (ii) \TTSB{} which is stable to arbitrary perturbations can occur in periodically-driven $D\geq 1$ open systems, such as the $\pi$-Toom model, whose dynamics effectively implement a form of error correction (Sec.~\ref{sec:toom}).
These two flavors of time crystals evade ergodicity by vastly different means, either by strongly localizing the degrees of freedom or by actively shrinking potentially ergodicity-generating fluctuations.
Between these two extremes lie two regimes of time-crystalline behavior that are exponentially long-lived: prethermal time crystals in closed systems (both classical and quantum) and activated time crystals in open systems.
Although ergodicity is deferred in both cases, it is inevitable in the long run~\cite{palmer1982broken,petersen1989ergodic,walters2000introduction}.

The possibility of \TTSB{} in driven open systems can be understood as one consequence of a more radical form of ergodicity breaking: fault-tolerant computation. 
There is a long history of understanding computation as a fundamentally physical process, and the subsequent constraints which arise from thermodynamics: ``Computers may be thought of as engines for transforming free energy into waste heat and mathematical work''~\cite{landauer1961irreversibility, bennett1982thermodynamics, wolpert2019stochastic}.
In this point of view, a time-crystal can be understood as a physical realization of a simple computer program: ``while true,  apply a global \textsf{NOT}-gate.''
If the program can execute perfectly despite faulty (noisy) gates, and with a physical implementation that relies only on local interactions, the execution of such a program can be understood as a non-equilibrium ``phase of matter,'' and the error threshold for fault-tolerance as a non-equilibrium phase transition into a time-crystalline phase.

The work of \cite{gacs1988simple, gacs2001reliable} has shown that fault-tolerant classical computation can indeed be realized as a  locally-interacting autonomous process in the thermodynamic limit, and this work was later used to prove that fault-tolerant quantum computing can be as well~\cite{harrington2004analysis, dauphinais2017fault}. One could thus realize a time-crystalline phase by repeatedly running the program ``\textsf{NOT}'' on such an error-corrected computer, either quantum or classical. 
In this sense, the existence of time-crystals in open systems is an elementary application  of  deeper results  regarding the physical realizability of  error-correction in autonomous, locally interacting systems. 

One open question is to understand how  the transition from exponentially to infinitely long-lived time crystals precisely occurs. Indeed, as disorder is increased in closed 1D quantum systems, one expects a sharp phase transition to occur
from the prethermal discrete time crystal (Sec.~\ref{prethermal}) to the Floquet-MBL discrete
time crystal (Sec.~\ref{MBLDTC}); one expects a similar transition from the activated DTC to the Toom-type time crystal, as one tunes a uniform anti-majority-vote rule toward the anti-majority NEC rule.

One possibility is that the exponentially long-lived time crystals' lifetimes --- e.g. scaling as $\sim A e^{{\omega_D}/J}$ in the prethermal case (Sec.~\ref{prethermal}) or as $\sim A e^{\Delta/T}$ in the activated case (recall that $\Delta$ is the quasi-energy barrier defined in Sec.~\ref{sec:activated_tc}) --- exhibit a diverging prefactor. 
For example, as the transition is approached, $A$ could diverge as $A\sim |g-{g_c}|^{-y}$, where $g$ is the tuning parameter.
An alternate possibility is that the effective energy scale $1/J$ or $\Delta$ might diverge
at the transition.

A second set of open questions is whether a direct transition is possible from a Floquet-MBL discrete time crystal to a $\pi$-Toom or G\'acs-type time crystal (or something continuously
connected to it). To the authors of this Colloquium, this possibility seems a bit  unlikely since
such a transition would have to occur precisely when
the coupling to a bath is zero, where the system
goes from closed to open.
On the other hand, a Floquet-MBL discrete time crystal is what we would create with a perfect quantum computer executing precisely the same set of gates at each time step.
Meanwhile, a $\pi$-Toom time crystal is what we might create with a noisy but fault-tolerant quantum computer in which error correction was governed by a probabilistic cellular automaton.
But since the former could be simulated by the latter, one might also logically expect them to be separated by no more than a second-order phase transition at which the rate of entropy production rises from zero with a discontinuous derivative.

\bibliography{DTC_rmp}

\end{document}